\def\x{{\bf x}}
\def\y{{\bf y}}
\def\p{{\bf p}}
\def\k{{\bf k}}
\def\j{{\bf j}}
\def\v{{\bf v}}
\def\betaL{\beta_{\rm L}}
\def\tr{{\rm tr}}
\def\SUtwo{SU(2)}
\def\SUtwoL{SU(2)$_{\rm L}$}
\def\SUtwoR{SU(2)$_{\rm R}$}

\def\taubig{\tau_{\rm big}}
\def\tausmall{\tau_{\rm small}}
\def\net{{\rm net}}
\def\mom{\pi}
\def\boldmom{\mbox{\boldmath$\pi$}}    
\def\nq{\rho}
\def\umklapp{{\rm umk}}
\def\t{t_{\rm M}}

\documentstyle [eqsecnum,preprint,aps,epsf,fixes]{revtex}
\tightenlines

\makeatletter           
\@floatstrue
\def\figure{\let\@capwidth\columnwidth\@float{figure}}
\let\endfigure\end@float
\@namedef{figure*}{\let\@capwidth\textwidth\@dblfloat{figure}}
\@namedef{endfigure*}{\end@dblfloat}
\makeatother

\begin {document}

\preprint {UW/PT-97-22}

\title  {Effective theories for real-time correlations in hot plasmas}

\author {Peter Arnold}
\address
     {%
     Department of Physics,
     University of Virginia,
     Charlottesville, VA 22901
     }%
\author {Laurence G. Yaffe}
\address
    {%
    Department of Physics,
    University of Washington,
    Seattle, Washington 98195
    }%
\date {September 1997}

\maketitle
\vskip -20pt

\begin {abstract}%
{%
{%
We discuss the sequence of effective theories needed to understand the
qualitative, and quantitative, behavior of real-time correlators
$\langle {\cal A}(t) {\cal A}(0)\rangle$ in ultra-relativistic plasmas.
We analyze in detail the case where $\cal A$ is a gauge-invariant
conserved current.
This case is of interest because it includes a correlation recently
measured in lattice simulations of classical, hot, \SUtwo-Higgs gauge
theory.
We find that simple perturbation theory, free kinetic theory,
linearized kinetic theory, and hydrodynamics are all needed to understand
the correlation for different ranges of time.
We emphasize how correlations generically have power-law decays
at very large times due to non-linear couplings to long-lived
hydrodynamic modes.
}%
}%
\end {abstract}


\section {Introduction}

There is a rich variety of spatial and temporal scales in hot,
weakly coupled, ultra-relativistic gauge theories, a selection
of which are shown in Table~\ref{tab:scales}.
Depending on the problem of interest,
a wide variety of methods can be appropriate for describing the
physics on different scales:
naive perturbation theory, hard thermal loops,
dimensional reduction, kinetic theory,
Vlasov equations, Langevin equations, hydrodynamics, and more.
For static properties of plasmas,
the tower of effective theories relevant to different distance scales
is well understood \cite{3d}.
These effective theories are all Euclidean quantum field theories,
in 3 or 4 dimensions, and may be systematically constructed using
standard renormalization group techniques.

For real-time properties, including the analysis of non-equilibrium response,
constructing an appropriate sequence of effective theories
which correctly disentangles dynamics on different distance
(or time) scales is much less straightforward.
Some of the difficulties are illustrated by the long confusion
over one of the basic time scales of hot non-Abelian plasmas:
the time scale for non-perturbatively large fluctuations
in gauge fields
(such as those responsible for baryon number violation in electroweak theory)
\cite {ASY}.
To elucidate the real-time dynamics of hot gauge theories,
one should understand the tower of effective theories appropriate
for different time scales and, at least in simple examples,
calculate the appropriate matching between such theories.

\begin {table}
    \begin {center}
    \tabcolsep 10pt
    
    \begin {tabular}{lll}
    \hline 
       typical particle wavelength               & $T^{-1\strut}$	&\\
       typical particle separation               & $T^{-1}$              &\\
       Debye length                              & $(gT)^{-1}$           &\\
       plasma frequency                          & $(gT)^{-1}$           &\\
       non-perturbative magnetic length scale	& $(g^2 T)^{-1}$
                                             & (non-Abelian theories)    \\
       mean free path: any-angle scattering      & $(g^2 T\ln)^{-1}$    &\\
       mean free path: large-angle scattering    & $(g^4 T\ln)^{-1}$    &\\
       temporal scale for topological transitions& $(g^4 T)^{-1}$
                                             & (non-Abelian theories)    \\[4pt]
    \hline
    \end {tabular}
    \end {center}
    \caption
        {%
           Orders of magnitude of various distance and time scales in hot,
           ultrarelativistic, gauge theory for weak coupling.
	   Here, and henceforth, ``$\ln$'' stands for $\ln(g^{-2})$,
	   except for the case of the mean free path for any-angle
	   scattering in an Abelian gauge theory, where it is a
	   genuine logarithmic infrared divergence.
	   (Naturally, $c \equiv \hbar \equiv 1$.)
        }
    \label {tab:scales}
    \medskip
\end {table}

The current interest in hot electroweak baryon number violation has
motivated substantial efforts in numerical simulations of hot, classical,
lattice gauge theory \cite {ambjorn,moore,tang&smit1,tang&smit}.
For certain time and distance scales, it has been argued that
classical simulations can adequately reproduce the behavior of
the underlying quantum plasma of interest
(or more generally, that certain classical lattice results can be
translated into corresponding results for the continuum quantum theory
\cite {Arnold}.)
Hence, these simulations are a source of ``data'' for
comparison with theoretical pictures of hot gauge theories.
In particular, there have been recent measurements
by Tang and Smit \cite{tang&smit}
of time-dependent correlations $\langle {\cal A}(t) {\cal A}(0) \rangle$
for several gauge invariant operators $\cal A$.
An interesting theoretical goal is
then to understand how best to describe the physics of, and calculate
the behavior of, such time-dependent correlators
for different ranges of time.

In order to keep the discussion focussed, we shall specialize to
the particular class of correlations where the operator ${\cal A}$ is the
spatial part $\j$ of a conserved gauge-invariant current $j_\mu$
(such as the electric charge current in QED,
or fermion number current in QCD).
As we shall explain below, one of the correlations
measured by Tang and Smit falls into this class.
We shall also focus on the spatial average of such currents,
\begin {equation}
   {\cal A}(t) = \j(t) \equiv V^{-1} \int d^3x \> \j({\x},t)
   ,
\end {equation}
as did Tang and Smit.
However, the principles and techniques underlying our discussion
are not limited to this chosen class of examples.
As our interest is in the dynamics of hot gauge theories, we
shall assume that the current $j_\mu$ involves particles which
feel a gauge force with coupling $g$.
We also assume that there is no chemical potential, so
that the total charge $\int d^3x \> j^0$ has zero expectation.

\begin {figure}
\vbox
   {%
   \vspace*{-0.2in}
   \begin {center}
      \leavevmode
      
      \epsfbox [150 45 500 350] {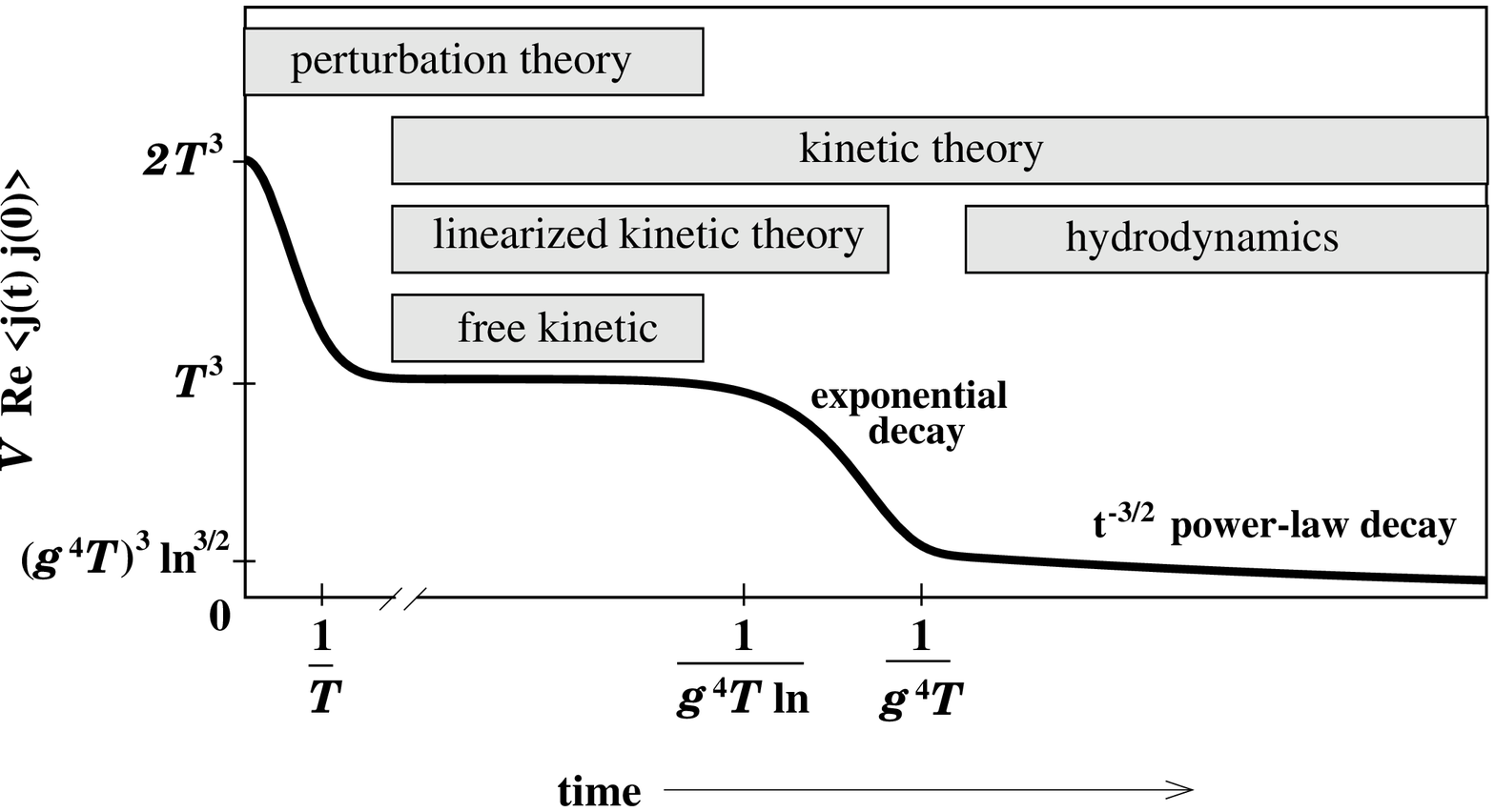}
   \end {center}
   \caption
       {%
         A qualitative picture of the real part of the current-current
         correlation $V\langle \j(t) \j(0) \rangle$ in a hot plasma,
	 for the case of bosonic charge carriers.
         The labels on the axes are only meant to denote orders of
         magnitude, except that the ``$2 T^3$'' mark is in fact twice the
         ``$T^3$'' mark in the small-coupling limit.
         The time axis should be thought of as linear
         before the break and logarithmic after; the vertical axis
         should be thought of as linear.
         $\ln$ is short for $\ln(g^{-2})$.
         The transient shown at $t \sim 1/T$ decays exponentially.
	 If the current is carried by fermions instead of scalars,
	 the resulting graph is identical except for the initial
	 transient, which starts from minus the ``$T^3$'' plateau value
	 instead of twice that value.
       \label{fig:summary}
       }%
   }%
\vspace{0.1in}
\vbox
   {%
   \begin {center}
      \leavevmode
      
      \epsfbox [150 45 500 390] {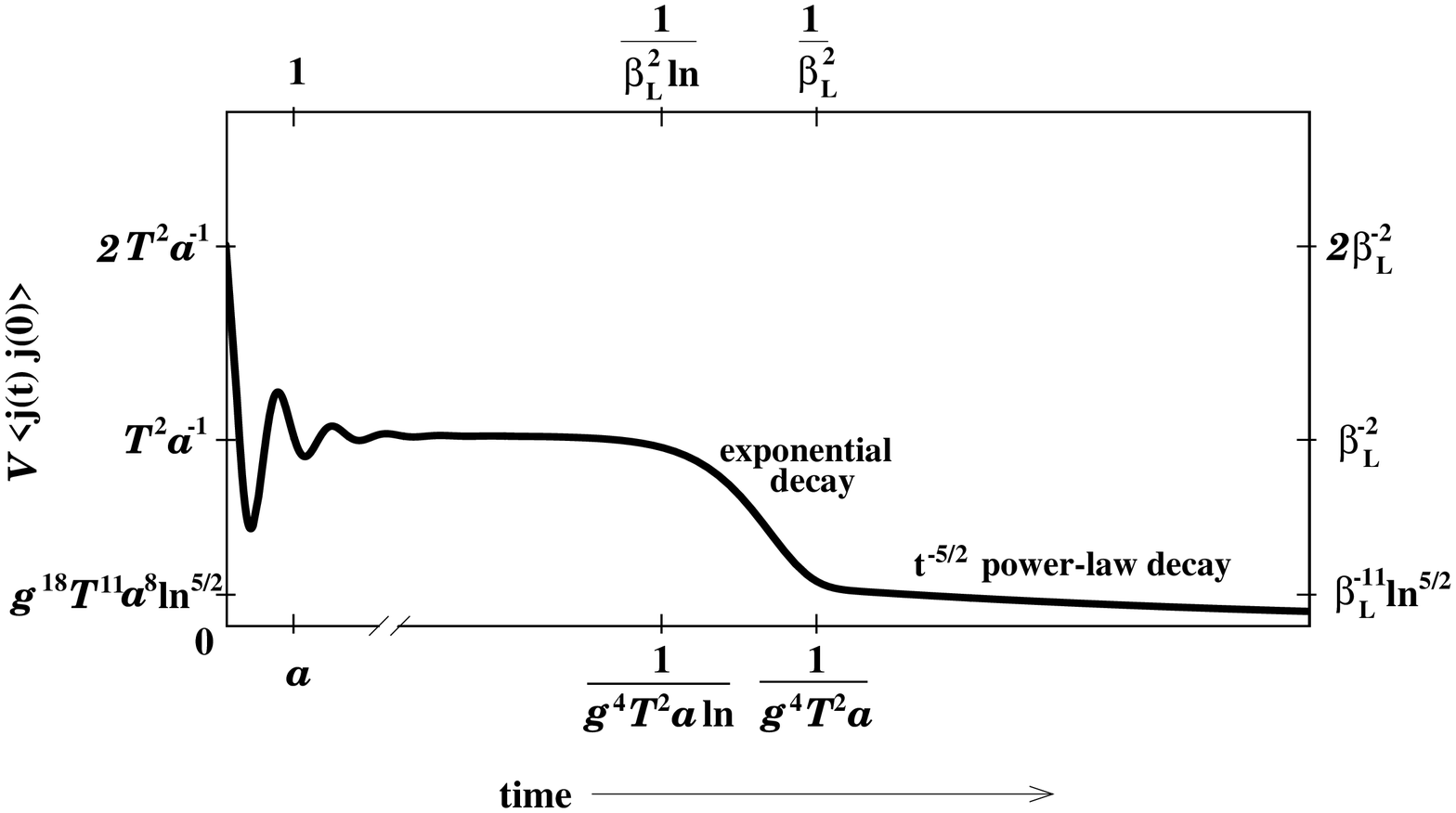}
   \end {center}
   \caption
       {%
         Similar to fig.~\protect\ref{fig:summary},
	 but for a classical field
         theory defined on an infinite spatial lattice with lattice spacing
         $a$.  We have specialized to a current $\j$ associated with a
	 non-Abelian symmetry.  The Abelian case is identical except that
         $g^{18} T^{11} a^8 \ln^{5/2}$ and $\betaL^{-11} \ln^{5/2}$
         on the left and right axes should be replaced by
         $g^{20} T^{12} a^9 \ln^{2}$ and $\betaL^{-12} \ln^{2}$,
	 respectively.
         The bottom and left axes are labeled in physical units.
         The top and right axis are labeled in lattice units, where
         the current $\j$ has been multiplied by $g^2$ as compared to
         the continuum normalization and $\betaL = \# (g^2 T a)^{-1}$.
         Here, $\ln$ is short for $\ln\betaL \sim \ln[(g^2 T a)^{-1}]$.
         The envelope of the transient oscillations
	 decays as $t^{-5/2}$.
       \label{fig:lattice}
       }%
   }%
\end {figure}

Fig.~\ref{fig:summary} qualitatively depicts our basic results for
the time dependence of the current-current correlator, and the physical
pictures best suited to understanding the dynamics at different time scales.
The purpose of this paper will be to explain this graph.
We have assumed that the gauge symmetry, and the symmetry corresponding
to the current, are unbroken.
We have also assumed that the current is normalized so that the
corresponding number operator $\int d^3x \> j^0$ counts each particle
as $\pm O(1)$ [rather than $\pm O(g)$, which would be conventional for the
electric current].
And we assume that the underlying theory is weakly coupled: $g^2(T)\ll 1$.

As will be discussed later, naive perturbation theory in the underlying
quantum field theory is sufficient to understand the behavior at early times.
At later times, it is appropriate to treat the excitations of the
theory as (quasi-)particles with definite position and momentum and to apply
kinetic theory.
At very late times, it is most convenient to focus on the long-wavelength,
collective, hydrodynamic excitations of the system.
As will be discussed below, the non-linear coupling of the current
to these hydrodynamic excitations gives rise to a long-time power-law
tail in the decay of the correlation function.

The behavior of the current-current correlation will turn out to be
dominated by excitations of hard thermal modes---that is, modes
with momentum $p$ of order $T$ in the continuum theory.
As a result, some of the interesting physics of hot plasmas
(such as the soft, $p \sim g^2 T$, \SUtwo-colored gauge fluctuations
responsible for electroweak baryon number violation) will not be probed
in our leading-order investigation, and the corresponding effective
theories \cite {ASY,Arnold,Huet&Son} do not appear
in fig.~\ref{fig:summary}.

Fig.~\ref{fig:lattice} shows the corresponding result
for a classical field theory on
an infinite spatial lattice.
(The finite volume case will be discussed in sec.~\ref{sec:finite volume}.)
We shall later explain the causes of the qualitative
differences ({\it e.g.}, the short time oscillations and the different
power-law decay of the long-time tail) from fig.~\ref{fig:summary}.
The differences between figs.~\ref{fig:summary}
and \ref{fig:lattice} reflect the fact that the current-current
correlator is sensitive to the hard thermal modes at all times,
even in the long time limit.
Such modes do not have occupation numbers large compared to one,
and will never be adequately described by a classical theory.
The differing origins of the long time tails in figs.~\ref{fig:summary}
and \ref{fig:lattice} will be discussed in detail in section
\ref{sec:tails}.

In this paper, we focus on a qualitative discussion of the correlator.
A quantitative treatment of some aspects will be deferred to a
companion paper \cite{companion}.
Throughout this paper, orders of magnitude given for time and distance
scales will refer to the case of continuum theory unless we
explicitly refer to classical lattice theory.
In sec.\ II, we examine the short-time
current correlations using simple perturbation theory,
and discuss the reduction to kinetic theory at times
large compared to $1/T$ (continuum) or $1/a$ (classical lattice).
A discussion of the leading-order perturbative result may also be found
in ref.~\cite{bodeker&laine},
which appeared as our work was being completed.
In sec.\ III, we discuss the current correlator in
the context of kinetic theory, and explain the
exponential decay at intermediate times, as shown in fig.~\ref{fig:summary}.
Section IV reviews the existence of long-time power-law tails in
systems with hydrodynamic behavior and shows how they arise in the
case of the current-current correlator.
In particular, we discuss the failure of
classical lattice simulations to reproduce the long-time
behavior of the continuum theory.
Sec.\ V explains how the behavior of the current
correlation changes qualitatively in finite volume systems.
Finally, having (we hope) made clear to the reader that
classical lattice simulations need not generate the
same long time behavior as high temperature quantum theories,
sec.\ VI briefly
reviews
the limited class of questions
for which real-time classical simulations may be useful.
The bulk of the paper focuses on currents in theories
for which the associated charge is carried by a scalar field.
(These are the theories for which it is interesting to compare
the dynamics of the continuum quantum theory with the corresponding
classical lattice field theory.)
However,
all the discussion of continuum quantum dynamics also applies
to theories where fermion fields carry the global charge of interest.
Appendix \ref{app:fermions} discusses the small differences
between results for bosonic and fermionic charge carriers.

\bigskip

In the remainder of this introduction, we review the correlator
measured by Tang and Smit.
Their interest was in electroweak theory,
and they simulated \SUtwo\ gauge theory coupled to a doublet Higgs scalar
$\phi$.
One of the correlations they measured was of the spatial average of
the gauge-invariant operator
\begin {equation}
   {\bf W}_a \equiv {1 \over 2i}\, \tr
       \left(
       [ \Phi^\dagger ({\bf D} \Phi) - ({\bf D}\Phi)^\dagger \Phi ]
       \tau_a
       \right)
   ,
\label {eq:W_a}
\end {equation}
where ${\bf D}$ is the \SUtwo\ covariant (spatial) derivative,
$\tau_a$ is a Pauli matrix,
and $\Phi$ is a $2\times2$ matrix made up of $\phi$ and its
charge conjugate:
\begin {equation}
   \Phi  =  ( i \tau_2 \phi^*,~ \phi )
   .
\end {equation}
They chose this operator because they were interested in studying the
thermal properties of gauge bosons.  Well into the symmetry-broken
phase of the theory, ${\bf W}_a$ is an operator that, to leading
order, simply creates or destroys massive gauge bosons:
${\bf W}_a \sim \langle\phi\rangle^2 {\bf A}_a$.
However, ${\bf W}_a$ has another interpretation: it is the
current associated with the global \SUtwoR\ custodial symmetry of
the gauge-Higgs system.  Specifically, the theory is invariant
under the \SUtwoL$\times$\SUtwoR\ symmetry
\begin {equation}
   \Phi \to L \Phi R^\dagger
   ,
\end {equation}
where $L$ and $R$ are independent \SUtwo\ rotations.
The \SUtwoL\ symmetry is gauged, and the \SUtwoR\
symmetry is an additional global symmetry for which ${\bf W}_a$
is the spatial part of the current.

The \SUtwoR\ symmetry is broken in the low temperature phase
of the theory (though the vectorial combination of \SUtwoR\
and \SUtwoL\ survives to guarantee equality of the W and
Z boson masses).
In this paper, we shall focus on the hot, symmetric phase of the
theory.
In this case, the operator ${\bf W}_a$ does {\it not} simply create
and destroy gauge bosons, and it is more usefully thought of as a current
carried by the scalars.
Indeed, ${\bf W}_a$ cannot couple to a state composed only of gauge
bosons because ${\bf W}_a$ carries an \SUtwoR\ charge
[$a$ is an \SUtwoR\ adjoint index]
while the gauge bosons do not.


\section {Perturbation theory}

In this and the following sections, we consider Abelian or non-Abelian
gauge theories with scalar and/or fermionic matter with a conserved,
gauge-invariant current $j_\mu$.
The simplest example to keep in mind is QED, coupled to either scalars
or fermions, where $j_\mu$ is the electric current.
For non-Abelian gauge theories, the gauge currents are not gauge-invariant,
and we do not wish to consider unphysical objects like gauge-dependent
correlators.
In the non-Abelian case, we must therefore
consider currents associated with additional
global symmetries among the matter fields,
such as the \SUtwoR\ custodial symmetry of the \SUtwo-Higgs theory.

\begin {figure}
\vbox
   {%
   \begin {center}
      \leavevmode
      
      \epsfbox {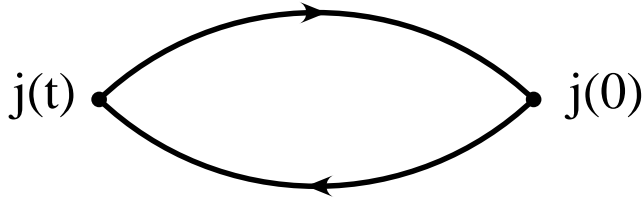}
   \end {center}
   \caption
       {%
          The leading-order perturbative diagram for
          $\langle\j(t) \j(0)\rangle$.
       \label{fig:pert}
       }%
   }%
\end {figure}

  One way to approximate correlators is by straightforward expansion
in diagrammatic perturbation theory.  Fig.~\ref{fig:pert} shows the leading
contribution to a current-current correlator $\langle \j(t) \j(0) \rangle$.
The result (for scalar fields) is
\begin {equation}
  V \langle j_i(t) j_j(0) \rangle =
   {\textstyle {1\over 3}} \,
  \delta_{ij} \,
  q^2
  \int_\p \> v_\p^2
  \left[\,
      n_{\omega_\p} e^{i \omega_\p t}
      + (1 + n_{\omega_\p}) e^{-i \omega_\p t}
  \,\right]^2
  ,
\label{eq:pert1}
\end {equation}
where $\omega_\p$ is the energy of an excitation with momentum $\p$,
$\v_\p \equiv \nabla_\p \omega_\p$ is its group velocity
(so $v_\p^2 = (p/\omega_p)^2$ for Lorentz invariant continuum theories),
and $\int_\p$ is short for $\int d^3p/(2\pi)^3$.
(See appendix \ref{app:fermions} for the case of fermionic charge carriers.)
$n_\omega$ is the Bose distribution,
\begin {equation}
   n_\omega \equiv {1\over e^{\hbar\beta\omega} - 1}
   ,
\end {equation}
where $\hbar$ has been shown explicitly for the first and last time in this
paper.
The case of classical field theory may be obtained by taking the
$\hbar \to 0$ limit of the Bose distribution, so that
\begin {equation}
   1+n_\omega \to n_\omega \to {1\over\beta\omega} \,.
\label{eq:classical}
\end {equation}
In (\ref {eq:pert1}), $q^2$ is the appropriate sum of squared charges,
for the particular current under consideration, of different particle species.
($q^2$ is 1 for the number current of a single complex scalar.
For non-Abelian currents such as (\ref{eq:W_a}), $q^2$ should be
multiplied by $2 \delta_{ab}$.)
For convenience, we will assume throughout that the currents of interest
are normalized such that $q^2$ is $O(1)$.

We can split the result (\ref{eq:pert1}) into an asymptotic constant $q^2 C$
and a decaying transient $q^2 D(t)$ by rewriting it as
\begin {mathletters}%
\label {eq:CD}%
\begin {equation}
  V \langle j_i(t) j_j(0) \rangle = \delta_{ij} \, q^2 [ \, C + D(t) ]
  ,
\end {equation}%
with
\begin {eqnarray}
  C &=& {\textstyle {2\over3}} \int_\p v_\p^2 \;
  n_{\omega_\p} (1 + n_{\omega_\p})
  ,
\label{eq:C}
\\
  D(t) &=&
    {\textstyle {1\over3}}
    \int_\p v_\p^2
    \left[
      n_{\omega_\p}^2 e^{2i\omega_\p t}
      + (1 + n_{\omega_\p})^2 e^{-2i\omega_\p t}
    \right]
  .
\label{eq:D}
\end {eqnarray}%
\end {mathletters}%
Note that $\langle \j(t) \j(0) \rangle$ is real for classical field
theory 
but not for quantum field theory, because
in the latter case $\j(t)$ and $\j(0)$ do not commute.  The imaginary
part only appears in the transient $q^2 D(t)$.

The asymptotic constant $q^2 C$ has a simple physical interpretation:
Eq.\ (\ref{eq:C}) is precisely the result one would obtain for the
expectation
$\langle \j(0) \j(0) \rangle$
in an ideal Bose gas of {\it particles} (and anti-particles)
rather than of fields.
These particles may be regarded as classical
(with definite position and momentum)
except that they satisfy Bose or Fermi statistics.
We shall refer to this as a kinetic theory description of the plasma.
The current is $q\v$ for each particle.
Because velocities of particles
do not change with time in an ideal gas,
$\langle \j(t) \j(0) \rangle = \langle \j(0) \j(0) \rangle
= q^2 \langle \v(0) \v(0) \rangle$ is
a constant for such a gas.
Note that $q^2 C$ is real, as it should be since $\j(t)$ is real and
commuting in the kinetic theory approximation.

The decaying transient $q^2 D(t)$, on the other hand, can be seen to
arise from an interference between particle and anti-particle contributions
in field theory.  Consider, for example,
a complex scalar field $\phi$ written in terms
of creation and annihilation operators:
\begin {equation}
   \phi(x) = \int_\p {e^{i\p\cdot\x} \over\sqrt{2\omega_\p}}
              \left(a_\p e^{i\omega_\p t} +
	      b_{-\p}^\dagger e^{-i\omega_\p t}\right)
              ,
\end {equation}
where we have used non-relativistic normalization for $a_\p$ and $b_\p$.
A (space-averaged) number current in a continuum scalar theory
would then be
\begin {equation}
   \j(t) = {-i \over V} \int_\x
            \phi^\dagger {\buildrel \leftrightarrow \over \nabla} \phi
   = \int_\p \v_\p \left[ N_\p + \overline N_{-\p}
          + a_\p^\dagger b_{-\p}^\dagger e^{-2i\omega_\p t}
          + b_{-\p}   a_\p   e^{ 2i\omega_\p t} \right]
   ,
\label {eq:j opr}
\end {equation}
where $N_\p = a_\p^\dagger a_\p$ and $\overline N_\p = b_\p^\dagger b_\p$
are the
number operators for particles and anti-particles, respectively
(and $\int_\x$ is, of course, short for $\int d^3x$).
Now consider the correlator $\langle \j(t) \j(0) \rangle$.
The non-oscillatory terms in the current (\ref{eq:j opr}) 
correspond to the kinetic theory description
and are responsible for
the asymptotic constant discussed before.  The oscillatory pieces
of the current, involving interference between particle and anti-particle
contributions, are responsible for the transient behavior of the correlator.
These oscillations become irrelevant at large times because
contributions from different momenta $\p$ decohere.

If we treat the particles as massless, then the only frequency scale in the
problem is $T$ for the continuum theory,
or the cutoff $a^{-1}$ for the classical lattice theory.
The time scale for the decay of the transient $D(t)$ is therefore
$1/T$ or $a$, respectively, and this is the origin of the transients at
those scales in figs.~\ref{fig:summary} and \ref{fig:lattice}.

In the case (\ref{eq:classical}) of classical field theory,
for a current carried by bosons,
the expression
(\ref{eq:CD}) for the transient contribution directly gives $D(0) = C$.
Hence, in leading-order perturbation theory,
the correlation decays by exactly a factor of 2 from its initial value,
as shown in fig.~\ref{fig:lattice}.
The same thing is true of the thermal contribution to the correlator
in quantum field theory.
The zero-temperature contribution to the correlator
[$n\to0$ in (\ref{eq:CD})], in the limit of negligible masses
(so that $\omega_p = |\p|$),
generates a $1/t^3$ imaginary part for the correlator.

The decay of the transient can be understood 
by analyzing the long-time behavior of D(t).%
\footnote{
   Long-time here means long compared to $1/T$ (or $a$) but applies only
   to what can be seen in the perturbative approximation of
   fig.~\ref{fig:pert}.
   It does not refer to the very long-time ($t \gtrsim 1/g^4T\ln$)
   physics shown in fig.~\ref{fig:summary}.
}
Expressing (\ref{eq:D}) in terms of a spectral weight,
\begin {eqnarray}
   D(t) &=&
   {\textstyle {1\over3}}
   \int_0^\infty d\omega \> \rho_{vv}(\omega)
    \left[
      n_{\omega}^2 \, e^{2i\omega t}
      + (1 + n_{\omega})^2 e^{-2i\omega t}
    \right]
\nonumber\\
   &=&
   {\textstyle {1\over3}}
   \int_{-\infty}^\infty d\omega \> 
       \rho_{vv}(\omega)
      \, n_{\omega}^2 \, e^{2i\omega t}
      ,
\label{eq:fourier}
\\
\noalign {\hbox {with}}
   \rho_{vv}(\omega) &=& \int_\p v_\p^2 \> \delta(\omega-\omega_\p)
   ,
\label{eq:rho}
\end {eqnarray}
converts the problem to a standard exercise analyzing the asymptotic
behavior of a Fourier transform.
(In (\ref{eq:fourier}), we used $1 + n_\omega = - n_{-\omega}$.)
This asymptotic behavior is determined by
non-analyticities of the integrand.
If we ignore the mass $m$, then in the continuum case with
$\omega_p = p$, the integrand is analytic in a strip about the real axis.
Consequently, the long-time behavior decreases exponentially with an
exponent determined by the imaginary part of the nearest singularity
in the complex $\omega$ plane (which is at $\pm 2\pi i T$).
So the decay shown in fig.~\ref{fig:summary} at $t \sim 1/T$ is
exponential.  In fact, in this particular case, one can find an
exact analytic result for all times, 
\begin {equation}
   C = {\textstyle {1\over 9}} T^3 , \qquad
   D(t) = {{\textstyle {1\over 3}} T^3 \over \sinh^2(2\pi Tt)} \left[
       (2\pi Tt + i\pi) \coth(2\pi Tt) - 1^{\vphantom{2}}
   \right] .
\label {eq:CD_massless}
\end {equation}

For a classical lattice theory, the situation is somewhat
different.  In this case,
(\ref{eq:classical}) and (\ref{eq:C})
then imply that the steady-state value $C$ is
dominated by momenta of order the inverse lattice spacing
$1/a$ rather than $T$, with the result that $C \sim T^2/a$.
Also, the density $\rho_{vv}(\omega)$ (\ref{eq:rho}) has
kink singularities generated from points at the edge of the Brillouin zone
where $\v_\p = 0$.
These are the well-known Van Hove singularities which
appear in the electronic density of states in crystals.
Such kink singularities produce an oscillating
power-law tail in Fourier transforms, with the frequency of oscillations
determined by the location of the singularities.
This is the source of the oscillations in the $t \lesssim a$ transient
shown in fig.~\ref{fig:lattice}.
The envelope of the oscillations decays like $t^{-5/2}$;
a detailed expression for the long-time behavior of $D(t)$ is
given in appendix~\ref{app:oscillate}.

Finally, let us go one step beyond the naive perturbation theory we
have considered so far.
It is well known that perturbation theory can be improved
for high temperatures by resumming ``hard thermal loops''
\cite {hard thermal loops}.
In particular, this means including the effective thermal mass in propagators.
The thermal contribution to the mass $m$ is $O(g T)$ in the continuum,
and $O(g \sqrt {T/a})$ 
for classical lattice theories.
In the presence of a mass, there will be discontinuities in
$\rho_{vv}(\omega)$ at $\omega=\pm m$.
These give rise to an additional power-law contribution
to the long-time behavior of $D(t)$ that oscillates with frequency $m$.
These oscillations have not been shown in figs.~\ref{fig:summary}
and \ref{fig:lattice}
because they are sub-leading compared to the total correlation:
the amplitude of the transient has already decayed to $O(g)$
by the time the mass oscillations begin.
See the complete expressions in appendix~\ref{app:oscillate}
for details.


\section {Kinetic Theory}
\label {sec:kinetic}

\begin {figure}
\vbox
   {%
   \begin {center}
      \leavevmode
      
      \epsfbox {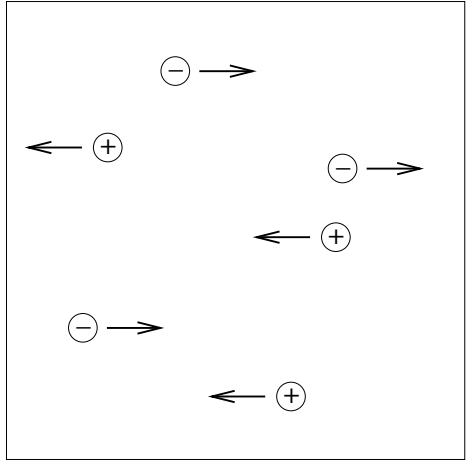}
   \end {center}
   \caption
       {%
         A pictorial example of a typical fluctuation with net current $\j$.
       \label{fig:fluctuation}
       }%
   }%
\end {figure}

In the last section, we saw that the leading-order perturbative result for the
correlator quickly settles down to the kinetic theory result for a
non-interacting gas.  In the kinetic theory picture, it is
easy to see that
the effect of interactions will become crucial at late times, and so simple
(one loop) perturbation theory must break down.%
\footnote
    {%
    The same conclusion can, of course, be reached from a diagrammatic
    analysis, albeit with more effort.
    For long times, one finds that contributions comparable in size to
    the lowest order diagram can be generated from an infinite series
    of ladder-like diagrams which contain (near) on-shell singularities.
    Summing these diagrams yields the same result which is obtained from
    an effective kinetic theory description.
    For a detailed analysis, in the context of a scalar theory,
    see \cite {jeon}.
    }
Currents are produced by fluctuations in the plasma.
Imagine focusing attention on some region of space (with a size
large compared to the inter-particle separation).
Suppose that, at time zero, a few more positive charged particles
happen to be moving left rather than right and/or a few more
negative charged ones are moving right rather than left,
as depicted in fig.~\ref{fig:fluctuation}, so that
there is a net current $\j(0)$ to the left.
As time progresses, the directions of the particles will be
progressively randomized by collisions, causing the net current to decay
and the correlation $\langle\j(t)\j(0)\rangle$ to approach zero.

A quantitative analysis of the effect of interactions will be
enormously simplified if one can indeed forget about quantum field theory
for times large compared to $1/a$ or $1/T$, and instead study the
dynamics using kinetic theory.
Kinetic theory applies to times and distances large compared
to particle energies and momenta.
In addition,
the validity of kinetic theory requires that de Broglie wavelengths
and scattering durations be small compared to the mean free path and
mean free time between collisions, respectively.
This ensures that particles can be viewed as propagating classically
(with on-shell energies) between collisions which are independent,
uncorrelated events.
Whether these conditions are satisfied depends on the
typical momenta and dominant scattering processes that
contribute to a given physical problem.
In the case at hand, it will be easiest to assume the
validity of kinetic theory and return later to check the conditions.


\subsection* {Mean free times}

\begin {figure}
\vbox
   {%
   \begin {center}
      \leavevmode
      
      \epsfbox {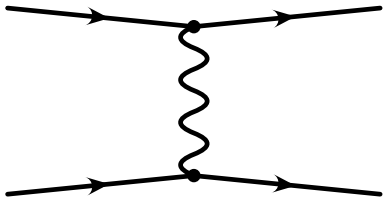}
   \end {center}
   \caption
       {%
          t-channel gauge boson exchange.  The solid lines represent any
          sort of particles that couple to the gauge force.
       \label{fig:tchannel}
       }%
   }%
\end {figure}

To begin, it will be convenient to review some well-known basic
scales associated with the times between collisions.
Scattering among particles that carry gauge charges
is dominated by $t$-channel exchange of a gauge boson, as depicted
in fig.~\ref{fig:tchannel}.
Consider a collision between typical particles in the
plasma, each with momentum of order $T$.
For continuum relativistic particles, the differential cross-section is
\begin {equation}
   n \, d\sigma  \sim  n g^4 {d\t \over \t^2}
                 \sim  {g^4 T} {d\theta \over \theta^3}
\label {eq:sigma}
\end {equation}
if we (temporarily) ignore plasma effects on the interaction.
Here, $\t=-Q^2$ is the virtuality of the exchanged gauge boson,
$\theta$ is the angle by which one of the particles is scattered,
and we've multiplied $d\sigma$ by the density $n \sim T^3$ of
scatterers.
The time between single, large-angle ($\theta \sim 1$) scatterings of
particles is therefore $O(1/g^4 T)$.

The total cross-section, however, is dominated by small-angle scatterings.
Eq.\ (\ref{eq:sigma}) looks like it will lead to a linear infrared
divergence in $t$, but plasma effects appear when $\sqrt{\t}$ is comparable to
the inverse Debye screening length $\sim gT$
(corresponding to $\theta \sim g$).
At this scale, Coulomb interactions are Debye screened, and
magnetic interactions are suppressed by Landau damping of
time-dependent, space-like magnetic fields.%
\footnote{
  At the technical level, one incorporates these effects by including the
  one-loop thermal self-energy in the propagator of the exchanged gluon.
}
The damping of such magnetic fields is a reflection of the
conductivity of the plasma
and Lenz's law that a conducting system resists changes in the
magnetic field.%
\footnote{
   We thank Guy Moore for pointing out this simple explanation.
}
If all Coulomb and magnetic interactions became completely irrelevant
for $\sqrt{\t} \ll gT$,
then the mean free time for scattering through any angle
would be order $1/g^2 T$ and dominated by $\theta \sim g$.
However, the plasma does not damp static magnetic fields, and so it does
not completely damp nearly-static ones.
Magnetic interactions remain significant
in the small region of phase space in which the energy transfer
(in the plasma rest frame) of the colliding particles is very small.
This phase space suppression turns out%
\footnote{
  For details see, for example, ref.~\cite{plasmon decay}.
}
to convert the linear small-$\t$ divergence of (\ref{eq:sigma})
into a logarithmic divergence $d\t/\t$ when $\sqrt{\t} \lesssim gT$.
In non-Abelian gauge theories, this logarithmic divergence is then
cut off by
non-perturbative magnetic physics
at a scale of $\sqrt{\t} \sim g^2 T$
($\theta \sim g^2$).
Hence, there is a logarithmic enhancement in the scattering rate
and so
a logarithmic suppression in the
mean free time
$\tausmall$
for any-angle scattering:
\begin {equation}
    \tausmall \sim {1 \over g^2 T\ln}
    ,
\end {equation}
where
``$\ln$'',
for non-Abelian theories,
means $\ln[(gT)/(g^2T)] \sim \ln(g^{-1})$.
The mean free time for any-angle scattering of gluons in particular
is typically referred to as the plasmon damping rate (in this case,
for hard gluons).

A different quantity of importance is the time it takes for the direction
of a single particle to be changed by a large angle ($\theta \sim 1$).
As discussed above, this can happen by a single, large-angle scattering
in a time of order $1/g^4 T$.
However, it can alternatively happen by a succession of
scatterings through a small angle $\Delta\theta$.  The time between
such scatterings is of order $(\Delta\theta)^2/g^4 T$, provided that
$\sqrt{\t} \gtrsim gT$ (or $\Delta\theta \gtrsim g$) so that there are
no suppressions by the screening and damping effects discussed above.
Treating successive
scattering as a random walk, order $1/(\Delta\theta)^2$ such scatterings
are needed to build up a total deviation of $\theta \sim 1$.  So a
succession of scatterings by $\Delta\theta$ {\it also} takes a time
of order $1/g^4 T$ to give large-angle scattering.  Since all possible decades
of $\Delta\theta$ contribute equally, the net result is a logarithmic
enhancement in the large-angle scattering rate,
$
   \taubig^{-1} \sim g^4 T \int {d(\Delta\theta) / (\Delta\theta)}
$.
The mean free time $\taubig$ for large-angle scattering is then
\begin {equation}
   \taubig \sim {1 \over g^4 T\ln}
   .
\end {equation}
This time, the
logarithm is cut off by $T$ in the ultraviolet and $gT$ in the infrared,
but it's still $\ln(g^{-1})$.
The non-perturbative physics associated with the scale $g^2 T$ in
non-Abelian plasmas is not involved at leading order.

For classical lattice theories, the above analysis can be repeated,
but the typical momenta of particles is $a^{-1}$ rather than $T$,
the density is $n \sim T a^{-2}$, the Debye mass scale is
$\betaL^{-1/2} a^{-1}$, and the scale for non-perturbative
magnetic physics in non-Abelian gauge theory is
$\betaL^{-1} a^{-1}$,
where the lattice coupling $\betaL^{-1}$ is order $g^2 T a$
in physical units.
This results in
$\taubig \sim \betaL^{-2} a/\ln(\betaL) \sim [g^4 T^2 a \ln(\betaL)]^{-1}$ for
the large-angle scattering time and
$\tausmall \sim \beta_L^{-1} a/\ln \sim (g^2 T^2 a \ln)^{-1}$ for the any-angle
scattering time,
where ``$\ln$'' is $\ln(\betaL)$ in non-Abelian gauge theory.


\subsection* {The correlator}

So which scattering time controls the decay of the
$\langle \j(t)\j(0) \rangle$ correlator?
Return to the picture
of a typical fluctuation at time zero
that has a net current $\j(0)$,
illustrated in fig.~\ref{fig:fluctuation}.
The total current $\j = \sum_s \j^{(s)}$ is the sum of currents
$\j^{(s)} = q^{(s)} \v^{(s)}$ of the individual particles (labeled by $s$).
Any one of those individual currents will only change substantially
when the velocity $\v^{(s)}$ changes substantially.%
\footnote{
  Recall that we have assumed that the current $\j$ is gauge-invariant,
  and so the corresponding symmetry commutes with the gauge interactions.
  As a result, the $q^{(s)}$ are not changed by small-angle scatterings
  due to gauge-boson exchange.
}
That means the relevant time scale for the randomization of the
current is the mean free time $\taubig$ for large-angle
scattering, rather than the mean free time $\tausmall$ for
any-angle scattering.  This is the physics responsible for the
period of exponential decay starting at time $1/(g^4 T \ln)$
in fig.~\ref{fig:summary}
[and time $1/(g^4 T^2 a\ln)$ in fig.~\ref{fig:lattice}].

There has been confusion in the literature on the above point.
In ref.~\cite{bodeker&laine}, for example,
it was suggested that the way to understand
the decay of the correlators measured by Tang and Smit is simply to
include thermal widths (the imaginary part of the self-energy) in the
internal propagators of the leading-order perturbative diagram,
fig.~\ref{fig:pert}.  The imaginary part of the self-energy measures the rate
for a particle to scatter into {\it any} other state and so includes
very small-angle scatterings---that is, the width is of order $\tausmall^{-1}$.
The inclusion of widths in the diagram of fig.~\ref{fig:pert}
makes the internal
propagators decay on a time scale of $\tausmall$ and therefore
{\it appears} to predict that
the correlation decays on a time scale of $\tausmall$
instead of $\taubig$.
The problem with this reasoning becomes apparent if one realizes that
it could be applied to any correlator at all, including the total
{\it charge} correlator $\langle j^0(t) j^0(0) \rangle$
(where $j^0(t) \equiv \int_\x j^0(\x,t)$).
By conservation of charge, the latter does not decay at all with time.
The problem is that the decay seemingly generated by resumming self-energies
into fig.~\ref{fig:pert} is canceled by other, higher-order diagrams in the
resummed perturbation theory.
An explicit example of such cancelations
for QED plasmas can be found in ref.~\cite{smilga}.
This highlights the dangers of using resummed perturbation theory
for intuition about physics at large times.
A (highly non-trivial) discussion of how kinetic theory may be seen to
arise from perturbative diagrams in scalar theory may be found
in ref.~\cite{jeon}.

We shall now outline how the rate of exponential decay
(for times of order $\taubig$)
can be calculated quantitatively from kinetic theory.
One starts with the Boltzmann equation for the probability
distributions $f(\x,\p;t)$ of particles (and anti-particles) in phase
space as a function of time.  The Boltzmann equation has the form
\begin {equation}
   (\partial_t + \v_\p \cdot \nabla_\x) f = C[f] \,.
\end {equation}
On the left is a convective time derivative of the distribution function.
The right hand side is a
collision integral representing the net rate at which particles are
scattered into a momentum state $\p$ minus the rate at which they are scattered
out.  For $2 \to 2$ scattering, it has the form
\begin {equation}
   C[f] = \int_{\p'\k\k'} |{\cal M}_{\p\p'\to\k\k'}|^2 \left[
      f_\p f_{\p'} (1 + f_\k) (1 + f_{\k'})
      - f_\k f_{\k'} (1 + f_\p) (1 + f_{\p'}) \right]
   ,
\end {equation}
where ${\cal M}$ is the scattering amplitude
(and indices labeling particle species are suppressed).

We are interested in the response in time of a fluctuation
away from equilibrium in the
(space-averaged) current
\begin {equation}
   \j(t) = V^{-1} \int_\x \int_\p q \, \v_\p \, f(\p,\x;t) \,.
\end {equation}
The response to fluctuations can be analyzed by linearizing $f$ about
the equilibrium distribution
$n(\omega_\p)$.%
\footnote{
   A classic discussion of most of these techniques can be found in
   chapter IV of ref.~\cite{physical kinetics}.
   See also ref.~\cite{heiselberg} for a somewhat related application of
   calculating the shear viscosity of hot QCD.
}
The resulting linearized Boltzmann equation has the form
\begin {equation}
   (\partial_t + \v_\p \cdot \nabla_\x)\,
   \delta f \approx {\cal C}[n] \, \delta f
   ,
\end {equation}
where ${\cal C}[n]$ is a linear integral operator acting on $\delta f$.
The decay of the space-averaged component of $\delta f$ is then
determined by the eigenvalues of ${\cal C}[n]$.
The eigenvector with smallest eigenvalue describes the most slowly
relaxing fluctuation (with a decay constant given by the eigenvalue),
and so dominates the long-time behavior in this approximation.
The operator ${\cal C}[n]$ has zero modes corresponding to
fluctuations in total charge, energy, and momentum,
but in the symmetry channel corresponding to the total spatial current
(C odd, CP even fluctuations)
the smallest eigenvalue of ${\cal C}[n]$ is non-zero.
In our companion paper \cite{companion},
we show how to calculate this decay exponent explicitly in leading-log
approximation.  The result is indeed $O(\taubig^{-1})$, as claimed
earlier.

Having realized that the relevant time scale is $O(\taubig)$ rather than
$O(\tausmall)$, we can now check the validity of kinetic theory.
We need to verify that the de Broglie wavelengths and scattering durations
are small compared to the time between these scatterings.
The fluctuations in the current are dominated by hard particles with
momentum $T$ and de Broglie wavelength $1/T$.
The dominant processes that contribute to the large-angle scattering
time $O(\taubig)$ are, as discussed earlier, scattering
events with momentum transfer between $gT$ and $T$.  The time between
such individual scatterings is $1/g^2 T$ and $1/g^4 T$, respectively,
and so we need kinetic theory to be valid for time scales as small as
$\Delta t \sim 1/g^2 T$.
The scattering duration for scattering with momentum transfer
$Q_\mu$ is order%
\footnote
    {%
    This estimate holds for typical scattering processes
    for which $q-Q_0$ and $q+Q_0$ are of comparable magnitude.
    For highly collinear scattering events, $|q-Q_0| \ll |q+Q_0|$,
    and the scattering time can be much larger than $1/gT$.
    However, the phase space of these exceptional scattering events
    is sufficiently restricted that they make a negligible
    contribution to $\taubig$ (or any of the physics under discussion).
    }
$1/|q-Q_0| \sim 1/\sqrt{Q^2} \lesssim 1/gT$.
This, and the de Broglie wavelength, are indeed small compared
to $\Delta t$ for weakly coupled theories, and so kinetic theory
is appropriate.%
\footnote{
  If the relevant time scale for the physics of interest
  had been $O(\tausmall)$ rather than $O(\taubig)$, then kinetic
  theory would be inappropriate (except at leading log order):
  it would break down for momentum
  transfers of order $g^2 T$ that contribute importantly to $O(\tausmall)$.
}

\subsection* {Large $N_f$ scaling}

    We digress momentarily to mention the scaling of the characteristic
time $\taubig$ for the decay of current correlations when there are a
large number $N_f$ of equivalent charged fields.
As usual,
the gauge coupling is assumed to scale as $1/N_f$ so that
$\lambda \equiv g^2 N_f$ is fixed as $N_f \to \infty$, and
the inverse Debye screening length $\sim \sqrt\lambda T$
has a finite large $N_f$ limit.
Repeating the earlier discussion, one finds that the mean free time
for large-angle scattering grows linearly with $N_f$,
\begin {equation}
    \taubig \sim {N_f \over \lambda^2 T \ln(\lambda^{-1})} \,.
\end {equation}
This means that there is a non-uniformity between
the large time and large $N_f$ limits,
with
$
    \lim_{N_f\to\infty} \lim_{t\to\infty} \langle \j(t) \j(0) \rangle = 0
$,
while
$
    \lim_{t\to\infty} \lim_{N_f\to\infty} \langle \j(t) \j(0) \rangle
    \ne
    0
$.
Consequently, calculations which use $1/N_f$ expansion techniques
\cite {large-N},
valid for arbitrary fixed times $t$, will not be able to see the
eventual decay of correlations occurring on time scales which grow as
$O(N_f)$.


\section {Long-time hydrodynamic tails}
\label {sec:tails}

\begin {figure}
\vbox
   {%
   \begin {center}
      \leavevmode
      
      \epsfbox {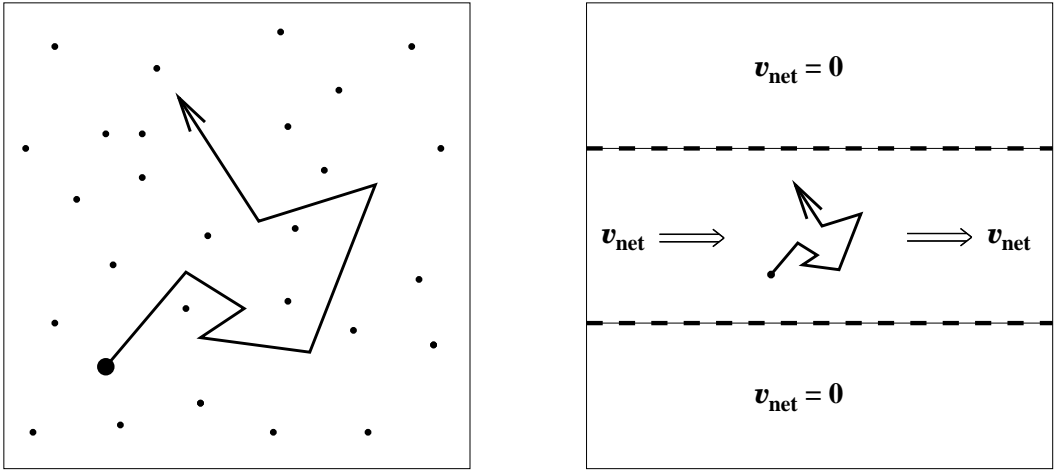}
   \end {center}
   \caption
       {%
          (a) The randomization of a particle's velocity by successive
          scatterings in the plasma.
          (b) The same process in the background of a large column
          of fluid moving with net velocity.
          In this case, the particle's velocity is not completely randomized
          but is biased by the net velocity $\v_\net$.
       \label{fig:randomize}
       }%
   }%
\end {figure}

In the last section, we argued that the velocities, and therefore the
currents, of individual particles get randomized on a time scale of
$\taubig$.  A picture of this randomization, as a result of successive
large-angle scatterings, is shown in fig.~\ref{fig:randomize}a.
Based on it, one might expect the current correlations
to show exponential decay at arbitrarily long times
(exactly as predicted by the linearized Boltzmann equation).
This prediction, however, is wrong, as was first discovered in 1970
in numerical simulations of a gas of classical billiard balls \cite{billiards}.
It was found that if one follows the velocity
$\v^{(s)}$ of a particular billiard ball $s$, the velocity auto-correlation
function $\langle\v^{(s)}(t) \v^{(s)}(0)\rangle$ has
{\it power-law} decay at large times.
Such power-law decay is a consequence of the presence of
hydrodynamic fluctuations with arbitrarily long relaxation times.%
\footnote{
   For a short discussion in the early literature, see
   ref.~\cite{tails}.
   For a review of the entire subject, see ref.~\cite{tail review}.
}
Fig.~\ref{fig:randomize}b illustrates the relevant physics.
Imagine a fluctuation in which there is a very long and wide column
of fluid that is in local equilibrium, but has a net average velocity $\v_\net$
with respect to the rest of the fluid.
After several large-angle scatterings, the velocity auto-correlation
$\langle \v^{(s)}(t) \v^{(s)}(0) \rangle$ of particles well inside
this column will not fall to zero, but will
approach the non-zero local equilibrium value
$\langle \v^{(s)} \rangle^2 = \v_\net^2$.
This remaining correlation will only decay on a time scale determined
by the diffusion of the particles out of the column of moving fluid.
(This includes two related effects: the particle $s$ under consideration
could leave the column, or the column's velocity could spread out and
degrade by the transverse diffusion of the other particles.
The latter process is parameterized by the shear viscosity of the fluid.)
Because diffusion is a random-walk process, the time it takes for
the remnant velocity correlation $\v_\net^2$ to decay is proportional
to the square of the transverse size $L$ of the column:
$t \sim (L/\bar v)^2/\taubig$,
where $\bar v$ is the typical particle speed.

Now that we've sketched the basic point, it is no longer necessary to
consider fluctuations of infinite spatial extent like the fluid column in
fig.~\ref{fig:randomize}b.
In general, after an arbitrarily long time $t \gg \taubig$,
the velocity auto-correlation of a given particle $s$ will be reduced to
roughly the square of the average velocity of the particles within
causal reach of particle $s$ by diffusion---that is, the particles
within a volume a radius $L \sim \bar v \sqrt{\taubig \, t}$.
The number of such particles is $N \sim n L^3$, and this average velocity
is then order $\bar v / \sqrt{N}$.
So, at late times,
\begin {equation}
   \langle \v^{(s)}(t) \cdot \v^{(s)}(0) \rangle
   \sim {\bar v^2 \over N}
   \sim \left[ n \bar v \, (\taubig \, t)^{3/2} \right]^{-1}
   \qquad\qquad
   (t \gg \taubig)
   .
\end {equation}
This $t^{-3/2}$ power-law decay is precisely what was measured in
simulations of billiard-ball gases.%
\footnote{
   In $d$ spatial dimensions, this becomes a $t^{-d/2}$ power law decay.
}

\begin {figure}
\vbox
   {%
   \begin {center}
      \leavevmode
      
      \epsfbox {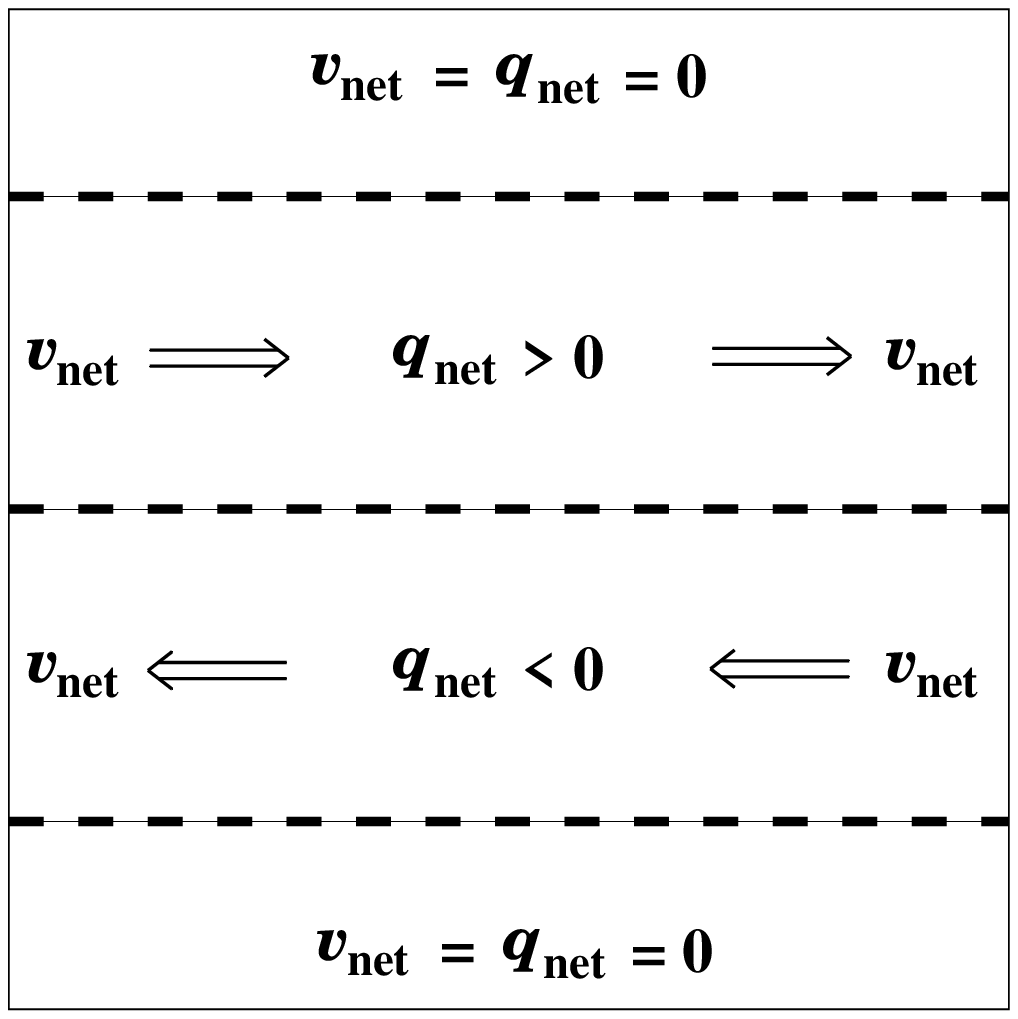}
   \end {center}
   \caption
       {%
          Coincident hydrodynamic fluctuations in both the velocity
          and charge density of the fluid,
	  giving rise to a long-lived fluctuation
          in the net current.
	  (Two columns are shown, with parallel currents
	  but opposite momentum and charge densities,
	  merely to emphasize that large scale current fluctuations
	  may occur with no change in total charge or momentum.)
       \label{fig:current}
       }%
   }%
\end {figure}

Now turn to our case of interest: the decay of the 
current-current correlation
$\langle \j(t) \j(0) \rangle$.
A long-wavelength fluctuation only in the velocity, as shown in
fig.~\ref{fig:randomize}b, is not enough to create a long-lived
net current.  We must additionally have a roughly coincident
long-wavelength fluctuation in the net charge, as shown in
fig.~\ref{fig:current}.  This is a long-lived fluctuation in
the current because, due to charge and momentum conservation,
neither the net local charge nor the net local
velocity can dissipate except by diffusion, and diffusive dissipation
takes arbitrarily long if the wavelength of the fluctuation is
arbitrarily long.  (This is in sharp contrast with the situation
shown in fig.~\ref{fig:fluctuation}, where there is a net current
but no net local velocity or charge.)
Because the current in fig.~\ref{fig:current} is the product of two small
fluctuations from equilibrium---the charge fluctuation and the
velocity fluctuation---it is a second-order effect in fluctuations
and will not be seen in a strictly linearized analysis of the problem.
This is why the linearized Boltzmann equation discussed in
section~\ref{sec:kinetic} does not yield a very-long-time power-law
tail.%
\footnote{
   It would be interesting to analyze in detail the Boltzmann equation
   treated to {\it second}-order in fluctuations, as it bears on our
   problem, and verify explicitly the time scale (discussed below) at
   which the linearized Boltzmann equation must break down, but we have
   not done so.
}

We will show in a moment how to use the equations of hydrodynamics to see
that the current correlator decays as $t^{-3/2}$, but first we give a
very rough and heuristic picture of this result.
Let $\bar j \sim \bar q \bar v$ be the typical absolute magnitude of the
current of an individual particle.  Then the total current at
time zero is of order $\bar j \sqrt{n V}$, where $n V$ is the total number
of particles in all of space.  This correctly reproduces
$V \langle \j(0) \j(0) \rangle \sim n \bar q^2 \bar v^2 \sim T^3$,
where $\j$ is the space-averaged current.
Now consider the $\langle \j(t) \j(0) \rangle$ correlation for times
large compared to $\taubig$.  In the case of the billiard-gas
velocity correlation,
all that was important was the average velocity of
a cell roughly of size $L \sim \bar v \sqrt{\taubig \, t}$.
For the current correlation,
what's important is both the average velocity and net charge of
such a cell.
These are of order $\bar v/\sqrt{n L^3}$ and $\bar q \sqrt{n L^3}$,
respectively, where $n L^3$ is the number of particles in the cell.
This gives a contribution to the current-current correlator of order
$\bar q \bar v$ for each cell.  Since there are
$N_{\rm cells} \sim V/L^3$ such
cells, each with randomly directed contributions to the correlation,
the total will be of order
\begin {equation}
   V \langle \j(t) \cdot \j(0) \rangle \sim
      {N_{\rm cells} \over V} \, (\bar q \bar v)^2
      \sim {\bar q^2 / \bar v \over (\taubig \, t)^{3/2}}
      \sim {1 \over (\taubig \, t)^{3/2}}
   \qquad
   (t \gg \taubig).
\label {eq:hydro decay}
\end {equation}

We can get a rough idea of when this power-law decay
takes over from the exponential decay discussed in the last section
by equating the correlator in the two approximations:
\begin {equation}
   T^3 e^{-t_{\rm cross}/\taubig}
   \sim {1\over (\taubig \, t_{\rm cross})^{3/2}}
\label{eq:crossing}
\end {equation}
implies
\begin {equation}
   t_{\rm cross} \sim \taubig \ln(\taubig T) \sim 1/(g^4 T) .
\end {equation}
At this point, the amplitude of the correlation is
\begin {equation}
   V \langle \j(t_{\rm cross}) \cdot \j(0) \rangle
   \sim (g^4 T)^3 \ln^{3/2}(g^{-1})
   ,
\end {equation}
as indicated in fig.~\ref{fig:summary}.


\subsection* {Hydrodynamic equations}

To be a little more systematic about the analysis of long-time tails,
we will now repeat our discussion within the context of the natural
effective theory for describing long-time excitations in plasmas:
hydrodynamics.  In general, hydrodynamics is the description of
long wavelength fluctuations of conserved densities,
{\it i.e.}, energy, momentum and charge densities.
Long wavelength fluctuations in these quantities can persist
an arbitrarily long time because they can relax only by diffusive processes
which transport the conserved quantities over long distances.
Consider a simple system with one conserved charge $q$ in addition to
the energy and momentum.
Let $\epsilon = T^{00}$, $\mom_i = T^{i0}$, and $\nq = \jmath^0$
be the energy, momentum, and $q$-charge densities.
Hydrodynamics consists of conservation laws,
\begin {mathletters}%
\label{eq:conserve}%
\begin {eqnarray}
   \partial_t \, \epsilon &=& -\nabla\cdot\boldmom ,
\label {eq:hydro eps}
\\
   \partial_t \, \mom_i    &=&  -\nabla_j \, T_{ij} ,
\label {eq:hydro mom}
\\
   \partial_t \, \nq      &=& -\nabla \cdot \j ,
\end {eqnarray}%
\end {mathletters}%
and constitutive relations which express the spatial currents in terms of
the conserved densities in the hydrodynamic limit.
In this example, we need constitutive relations for the stress tensor
$T_{ij}$ and the spatial charge current $\j$.
The constitutive relations may be parameterized by writing down, in
a small-momentum expansion, all possible terms consistent with symmetries
(including boost invariance) of the resulting equations.
Typically, one expands only to first-order in momentum.
For our purposes, it will be least cumbersome to also consider an
expansion in fluctuations $\delta\epsilon$, $\boldmom$,
and $\nq$ away from global equilibrium.%
\footnote{
  As everywhere in this paper, we are assuming that there is no
  chemical potential(s), hence the equilibrium change
  density is zero.
  We also choose to work in the rest frame of the fluid,
  so that $\boldmom$ vanishes in equilibrium.
}
Then, for example, terms allowed by symmetry for linearized hydrodynamics are
\begin {mathletters}%
\begin {eqnarray}
   T_{ij} &=& \delta_{ij} \left(
     a + b \, \delta\epsilon - \gamma_\zeta \nabla\cdot\boldmom \right)
     - \gamma_\eta \left( \nabla_i \mom_j + \nabla_j \mom_i
          - {\textstyle{2\over3}} \delta_{ij}\nabla\cdot\boldmom \right) ,
\label {eq:hydro stress}
\\
   \j &=& -D_q \nabla \nq \,,
\end {eqnarray}%
\label{eq:jconstituitive1}%
\end {mathletters}%
to first order in gradients.
The constants are usually expressed in more physical terms as
$a = P$, $b=v_{\rm s}^2$, $\gamma_\eta = \eta/\langle\epsilon+P\rangle$,
and $\gamma_\zeta = \zeta/\langle\epsilon+P\rangle$, 
where $P$ is the pressure, $v_{\rm s}$ is the speed of sound,
$\eta$ and $\zeta$ are the shear and bulk viscosity,
and $\langle\epsilon+P\rangle$ is the equilibrium enthalpy density.
$D_q$ is the diffusion constant for the conserved charge $q$.

In the linearized approximation, the space-averaged current is
automatically zero:
\begin {equation}
   \j(t) \approx -\int_\x D_q \nabla \nq(\x, t) = 0 \,.
\end {equation}
To study correlations of the space-averaged current, we will
therefore need to consider the next higher order terms in the
current constitutive relations (\ref{eq:jconstituitive1}).
Incorporating the next term consistent with symmetries%
\footnote{
   Including charge conjugation,
   under which $j_\mu \to -j_\mu^*$.
}
produces,
\begin {equation}
   \j = -D_q \nabla \nq + c \nq \, \boldmom ~~+~~ (\hbox{higher order})
   .
\end {equation}
In fact, the constant $c$ is determined by boost invariance, and
this expansion is just
\begin {eqnarray}
   \j &=& -D_q \nabla \nq + \nq \, \v ~~+~~ (\hbox{higher-order})
   ,
\label{eq:jconstituitive2}
\\\noalign{\hbox {where}}
   \v &=& {\boldmom \over \langle \epsilon + P\rangle}
            ~~+~~ (\hbox{higher-order})
\end {eqnarray}
is the local fluid velocity.
This has been a rather roundabout way to realize that there is a contribution
of the form $\nq\v$ to the current, but this methodical approach
will be useful when we move on to discuss long-time tails in lattice theories.

We can now analyze the hydrodynamic decay of current correlations by
using the second-order constitutive relation (\ref{eq:jconstituitive2})
for $\j$ to express
\begin {eqnarray}
   \langle \j(t) \cdot \j(0) \rangle
   &\simeq&
     \langle \int_\x \nq(\x,t) \v(\x,t)
             \cdot \int_\y \nq(\y,0) \v(\y,0) \rangle
\nonumber\\
   &=& \langle \int_\k \tilde \nq(\k,t) \tilde\v(-\k,t) \cdot
          \int_{\k'} \tilde \nq(-\k',t) \tilde\v(\k',0) \rangle ,
\end {eqnarray}
where $\tilde\nq (\k,t)$ denotes the spatial Fourier transform 
for $\nq(\x,t)$, {\em etc}.
One may now use linearized hydrodynamics to
compute the correlation (which factorizes at this order),
\begin {equation}
   \langle \j(t) \cdot \j(0) \rangle
   \simeq \int_{\k\k'}
        \langle \tilde \nq(\k,t) \tilde \nq(-\k',0) \rangle \>
        \langle \tilde\v(-\k,t) \cdot \tilde\v(\k',0) \rangle .
\label{eq:moo}
\end {equation}
Solutions to the hydrodynamic equations tell one about the decay of
correlations.
The linearized $\nq$ equation,
$\partial \nq / \partial t = D_q \nabla^2 \nq$,
gives
\begin {equation}
   \tilde \nq(\k,t) \simeq \tilde \nq(\k,0) \, e^{-D_q k^2 t}
   ,
\label {eq:linear charge}
\end {equation}
which should be interpreted in the present context as meaning
\begin {equation}
   \langle \tilde \nq(\k,t) \tilde \nq(-\k',0) \rangle
   \simeq \langle \tilde \nq(\k,0) \tilde \nq(-\k',0) \rangle \,e^{-D_q k^2 t}
   .
\end {equation}
To find the equal-time fluctuation spectrum
$\langle \tilde \nq(\k,0) \tilde \nq(-\k',0) \rangle$,
we have to step back from hydrodynamics and realize that in position
space the charge density $\nq$ is only correlated over very small distances
(of order $1/T$).
As far as the long distance scales of hydrodynamics are concerned,
this correlation can be approximated as local:
\begin {equation}
   \langle \nq(\x,0) \nq(\y,0) \rangle \simeq \chi_q \, \delta^{(3)}(\x - \y)
   ,
\end {equation}
where $\chi_q$ is the charge susceptibility,
\begin {equation}
   \chi_q \equiv \int_\x \langle \nq(\x,0) \nq({\bf 0},0) \rangle
   .
\label {eq:chin}
\end {equation}
Consequently,
\begin {equation}
   \langle \tilde \nq(\k,t) \tilde \nq(-\k',0) \rangle
      \simeq \chi_q \, \delta^{(3)}(\k-\k') \, e^{-D_q k^2 t} \,.
\end {equation}
The corresponding solutions for the behavior of velocity $\tilde\v(\k,t)$
(or momentum density) fluctuations
split into the two cases of transverse ($\tilde\v \perp \k$)
and longitudinal ($\tilde\v \parallel \k$) perturbations.
The solutions in the two cases behave as
\begin {mathletters}%
\begin {eqnarray}
   \tilde \v_\perp(\k,t) &\simeq&
   \tilde \v_\perp(\k,0) \, e^{-\gamma_\eta k^2 t} ,
\label {eq:transverse}
\\
   \tilde \v_\parallel(\k,t) &\simeq&
   \tilde\v_\parallel(\k,0) \, \cos(v_{\rm s} k t) e^{-\gamma_{\rm s} k^2 t/2}
   -
   {iv_{\rm s} \, \hat \k\over \langle \epsilon{+}P\rangle} \,
   \delta\tilde\epsilon(\k,0) \,
   \sin(v_{\rm s} k t) \, e^{-\gamma_{\rm s} k^2 t/2}
\label {eq:parallel}
\end {eqnarray}%
\label {eq:linear velocity}%
\end {mathletters}%
Transverse fluctuations (\ref {eq:transverse}) relax diffusively at
a rate controlled by the shear viscosity,
while longitudinal fluctuations (\ref {eq:parallel})
are propagating sound waves which decay at a rate
governed by the sound attenuation constant
$\gamma_{\rm s} = {4\over3} \gamma_\eta + \gamma_\zeta$.
The net result for the current-current correlator (\ref{eq:moo}) is
\begin {equation}
   V \langle{\j(t)\cdot\j(0)\rangle} \simeq \chi_q \chi_v \left[\>
      2 \int_\k e^{-(D_q+\gamma_\eta) k^2 t}
      + \int_\k \cos(v_{\rm s} k t) \, e^{-(D_q+\gamma_{\rm s}/2) k^2 t}
   \,\right]
   ,
\end {equation}
where the susceptibility $\chi_v$ is
\begin {equation}
   \chi_v \equiv {1\over3 \langle\epsilon+P\rangle^2}
      \int_\x \langle \boldmom(\x,0) \cdot \boldmom({\bf 0},0) \rangle
   .
\end {equation} 
The first integral gives the $t^{-3/2}$ behavior discussed previously.
The second integral decays exponentially in time due to decoherence
in the spectrum of sound waves.
The long-time result is then
\begin {equation}
   V \langle{\j(t)\cdot\j(0)\rangle} \simeq
   {2 \chi_q \chi_v \over \left[4\pi(D_q+\gamma_\eta) \, t\, \right]^{3/2}}
   .
\end {equation}
The susceptibilities $\chi_q$ and $\chi_v$ are equal-time quantities and
are easily computed using perturbation theory in the original quantum
field theory.  They are order $T^3$ and $T^{-3}$ respectively.
The constants $D_q$ and $\gamma_\eta$ are related to diffusion and
are $O(\taubig)$.
They can be evaluated precisely in any given theory from the
linearized Boltzmann equation.
Techniques for calculating such coefficients, under the simplifying
assumption that $\ln(g^{-1}) \gg 1$, can be found in
ref.~\cite{heiselberg},
which computes the shear viscosity $\eta$ (and so $\gamma_\eta$)
for QCD in this approximation.
In any case, the order of magnitude of the correlator at long times is
\begin {equation}
   V \langle{\j(t)\cdot\j(0)\rangle} \sim {1\over(\taubig \, t)^{3/2}}
   ,
\label {eq:hydro decay2}
\end {equation}
in agreement with our previous estimate (\ref{eq:hydro decay}).


\subsection* {Not-quite-hydrodynamics on the lattice}

Now let's try the same analysis for theories on a spatial lattice.
Because the lattice breaks continuous translation invariance,
there is one crucial difference between the lattice and continuum theories:
momentum is not an extensive, additively conserved quantity in the lattice
theory.
More specifically, momentum is only conserved modulo $2\pi/a$ on a
(simple cubic) lattice.
In the continuum, boosting an equilibrium state yields another equilibrium
state: A uniformly boosted plasma, with non-zero momentum
density $\boldmom$ and total momentum $\boldmom V$,
is a stable equilibrium state.
 On the lattice, however, the total momentum $\boldmom V$ will decay away
due to microscopic $2\to2$ particle collisions%
\footnote{
   Depending on the details of the lattice Hamiltonian, $2\to1$ and
   $1\to2$ processes may also be possible, in which case the time scale
   discussed below will be even shorter.  For the canonical
   Kogut-Susskind Hamiltonian, however, $2\to1$ and $1\to2$ processes
   are forbidden by energy and lattice momentum conservation.
}
that violate momentum by
multiples of $2\pi/a$ in one or more lattice directions.
(These are just the umklapp processes of solid state physics.)
This $2\to2$ process is dominated by momenta, and momentum transfers,
of $O(1/a)$.
The corresponding time scale is
the same as that for individual large-angle scatterings:
$1/O(g^4 T^2 a)$.%
\footnote{
  It would be convenient to work in lattice units as then everything
  can be expressed in terms of the single variable $\betaL \sim (g^2 a T)^{-1}$
  instead of $g$, $a$, and $T$ separately.  In lattice units,
  the current should be scaled with an extra power of $g^2$, so that
  $j_{\rm lat}(t_{\rm lat}) \sim g^2 a^{-3} j(a t_{\rm lat})$.
  However, we will stick to physical
  units to facilitate comparison with previous continuum results.
  In such comparisons, one should think of $1/a$ as being order $T$, since
  these are the UV momentum scales in the two cases.
}
As a result, large-scale fluctuations in momentum may decay
on the same time scale as other microscopic scattering processes;
consequently, momentum density should no longer be treated as one
of the fundamental independent variables for ``hydrodynamics'' on
the lattice.

So what does the appropriate long-distance and long-time theory look like?
The conservation laws (\ref{eq:conserve}) are now reduced to
\begin {mathletters}%
\begin {eqnarray}
   \partial_t \, \epsilon &=& -\nabla\cdot\boldmom ,
\\
   \partial_t \, \nq      &=& -\nabla \cdot \j ,
\end {eqnarray}%
\end {mathletters}%
and we now need an effective constituent equation for the momentum
density $\boldmom$, as well as $\j$, in terms of just
energy and charge densities.
The lowest-order possibility
for $\boldmom$ is
\begin {equation}
   \boldmom = -D_\epsilon \nabla\epsilon ~~+~~ (\hbox{higher order})
   .
\end {equation}
The lowest-order possibility for the current $\j(\x,t)$ that will
give a non-vanishing contribution to the spatial average of $\j$ is
\begin {equation}
   \j = -D_q \nabla \nq + c_1 \, \delta\epsilon \nabla \nq
                     + c_2 [\nq,\nabla \nq]
        ~~+~~ (\hbox{higher order})
   .
\label{eq:jlattice}
\end {equation}
The commutator term $[\nq,\nabla \nq]$ appears only if the current $\j$ is
associated with a non-Abelian global symmetry
(as is the case for Tang and Smit's $\bf W$ correlator).

Following the procedure we used for the continuum case, the long-time
tail of the current correlator is then given by
\begin {eqnarray}
   V \langle{\j(t)\cdot\j(0)\rangle}
   &\simeq&
      c_1^2 \, \chi_\epsilon \chi_q \int_\k k^2 e^{-(D_\epsilon + D_q) k^2 t}
      + c_2^2 \, \chi_q^2 \int_\k k^2 e^{-2 D_q k^2 t}
\nonumber\\
   &=& {6 \pi\, c_1^2 \, \chi_\epsilon \chi_q
         \over [4\pi(D_\epsilon + D_q) \, t \,]^{5/2}}
     + {6 \pi\, c_2^2 \, \chi_q^2
         \over [8\pi D_q \, t \,]^{5/2}}
   ,
\label{eq:lat tail}
\end {eqnarray}
where the susceptibility $\chi_\epsilon$ is defined analogously to $\chi_q$
(\ref{eq:chin}), with $\nq \to \delta\epsilon$,
and is better known as $T^2$ times the specific heat
(more precisely, $T^2$ times
the heat capacity per unit volume).

As in the continuum quantum theory,
the susceptibilities can be computed in perturbation theory,
and the diffusion coefficients could be calculated from the linearized
Boltzmann equation.
The susceptibilities are of order
\begin {equation}
   \chi_q \sim T^2 a^{-1} ,
   \qquad
   \chi_\epsilon \sim T^2 a^{-3} ,
\end {equation}
and the diffusion constants are
\begin {equation}
   D_q \sim \sim \taubig \sim (g^4 T^2 a \ln)^{-1} ,
   \qquad
   D_\epsilon \sim (g^4 T^2 a)^{-1}
   .
\end {equation}
$D_\epsilon$ does not have a log suppression because the diffusion of
energy density is controlled by umklapp processes which, as asserted
earlier, are dominated by large-momentum transfer processes.%
\footnote{
  One way to understand how umklapp processes determine $D_\epsilon$ is
  to return to the continuum case of eqs.\ (\protect\ref{eq:hydro mom}) and
  (\protect\ref{eq:hydro stress}).  Schematically,
  $\dot{\boldmom} \sim - v_s^2 \nabla\epsilon + \gamma \nabla \nabla \boldmom$.
  Now add the effect of momentum decay on the lattice with a time
  scale $t_\umklapp$ to get, schematically,
  $\dot{\boldmom} \sim - v_s^2 \nabla\epsilon + \gamma \nabla \nabla \boldmom
     - t_\umklapp^{-1} \boldmom$.
  In the hydrodynamic limit of arbitrarily small frequencies and
  momentum, the $\dot{\boldmom}$ and $\nabla \nabla\boldmom$ terms can be ignored,
  giving $\boldmom \sim - v_s^2 t_\umklapp \nabla \epsilon$.  Hence
  $D_\epsilon \sim v_s^2 t_\umklapp$.
}
(Log suppressions in other diffusion coefficients came, in
contrast, from small-momentum transfers---that is, from small-angle
scattering.)

\begin {figure}
\vbox
   {%
   \begin {center}
      \leavevmode
      
      \epsfbox {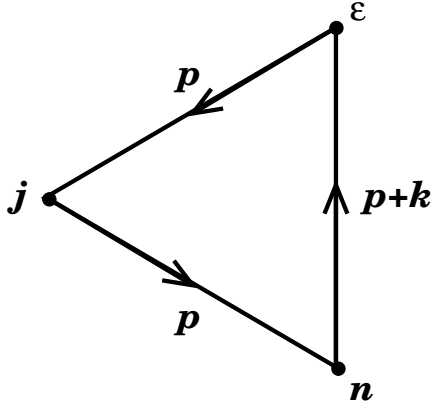}
   \end {center}
   \caption
       {%
          Leading-order diagram for computing the
          $\langle \j \epsilon \nq \rangle$ correlation.
       \label{fig:jen}
       }%
   }%
\end {figure}

The final ingredients are the coefficients
$c_1$ and $c_2$ of the constitutive relation (\ref{eq:jlattice}).
Physically, these characterize how fluctuations in $\j$ correlate with
fluctuations in $\epsilon$ and $\nq$.
For simplicity, let's assume the symmetry associated with $\j$ is
Abelian and ignore $c_2$.
We can then compute $c_1$ from a ratio of equal-time correlators,
\begin {eqnarray}
   c_1 &=& V^2 \; \lim_{\k\to0}
     { i\k \cdot \langle \j(0) \tilde\epsilon(\k,0) \tilde \nq(-\k,0) \rangle
       \over
       \k^2 \langle |\tilde\epsilon(\k,0) \tilde \nq(-\k,0)|^2 \rangle }
\label {eq:c1}
\\
   &\simeq& {V^2\over\chi_q\chi_\epsilon} \>
     \lim_{\k\to0}
     {i\k \over \k^2} \cdot
     \langle \j(0) \tilde\epsilon(\k,0) \tilde \nq(-\k,0) \rangle
   ,
\end {eqnarray}
as may be verified by substituting (\ref{eq:jlattice}) into
(\ref{eq:c1}).
The equal-time three-point correlator can be computed,
with resummed perturbation theory
in the original field theory, from the diagram shown in fig.~\ref{fig:jen}.
This diagram is dominated by momenta of order the thermal mass
$m \sim g \sqrt {T/a}$ and yields%
\footnote{
  When working in physical units, an easy way to get the order of a
  diagram in the classical theory is to include a power of $T$ for
  each propagator and then make up dimensions with $m$ if the diagram
  is infrared dominated or $a^{-1}$ if UV dominated.
}
\begin {equation}
   c_1 \sim {a^4\over T} \ln\!\left(1\over ma\right)
   .
\end {equation}
The order of the long-time tail (\ref{eq:lat tail}) is then
\begin {equation}
   V \langle{\j(t)\cdot\j(0)\rangle}
   \sim
   {T^2 a^4 \ln^2\!\left({1\over ma}\right) \over
   [(D_\epsilon + D_q)\, t \,]^{5/2}}
   .
\end {equation}
The cross-over between exponential decay
and the hydrodynamic tail occurs at
\begin {equation}
   T^2 a^{-1} e^{-t_{\rm cross}/\taubig} \sim
   {T^2 a^4 \ln^2\!\left(1\over ma\right) \over
   [(D_\epsilon + D_q)t_{\rm cross}]^{5/2}}
   ,
\end {equation}
analogous to (\ref{eq:crossing}).  This time is
\begin {equation}
   t_{\rm cross} \sim 1/(g^4 T^2 a)
   ,
\end {equation}
at which point the amplitude is
\begin {mathletters}%
\begin {eqnarray}
   V \langle \j(t_{\rm cross}) \cdot \j(0) \rangle
   &\sim& g^{20} \, T^{12} \, a^{9} \, \ln^{2}
   \qquad\qquad
   (\hbox{Abelian symmetry})
\\
   &\sim& {1 \over g^4 a^3} \> \betaL^{-12} \ln^{2}(\betaL) \,.
\end {eqnarray}%
\label {eq:lat-abelian-tails}%
\end {mathletters}%
The scaling of this amplitude like the {\it tenth} power of $\betaL^{-1}$
relative to the $t=0$ correlation (ignoring the logs)
means that the hydrodynamic tail does not appear on a fine
lattice ($\betaL \ll 1$) until the correlation is very small.

For a non-Abelian $\j$, we need to consider the contributions to the
hydrodynamic tail from $c_2$ as well.  These may be worked out similarly
to the $c_1$ contributions above and also give a $t^{-5/2}$ tail.
However, the $c_2$ contributions give a much larger amplitude for
the tail.
One can easily find that
\begin {mathletters}%
\begin {eqnarray}
   c_2 &=&
   -{\textstyle {i \over 2}} V^2
   \; \lim_{\k\to0}
     { \k \cdot \langle {\rm tr} \> \j(0) [ \tilde\nq(\k,0), \tilde \nq(-\k,0)] \rangle
       \over
       \k^2 \> \langle {\rm tr} \> [\tilde\nq(\k,0), \tilde \nq(-\k,0)]^2 \rangle }
\\
   &\sim& {a^2 \over mT} ,
\end {eqnarray}%
\end {mathletters}%
and
\begin {mathletters}%
\begin {eqnarray}
   V \langle \j(t_{\rm cross}) \cdot \j(0) \rangle
   &\sim& g^{18} \, T^{11} \, a^{8} \, \ln^{5/2}
   \qquad\qquad
   (\hbox{non-Abelian symmetry})
\\
   &\sim& {1 \over g^4 a^3} \> \betaL^{-11} \ln^{5/2}(\betaL)
   ,
\end {eqnarray}
\label {eq:lat-nonabelian-tails}%
\end {mathletters}%
as indicated in fig.~\ref{fig:lattice}.


\section {Finite Volume}
\label{sec:finite volume}

So far, we have only considered the behavior of the correlator in
infinite volume.
Real lattice simulations are, of course, done in finite volume.
How does this change things?

\begin {figure}
\vbox
   {%
   \begin {center}
      \leavevmode
      
      \epsfbox [100 30 450 350] {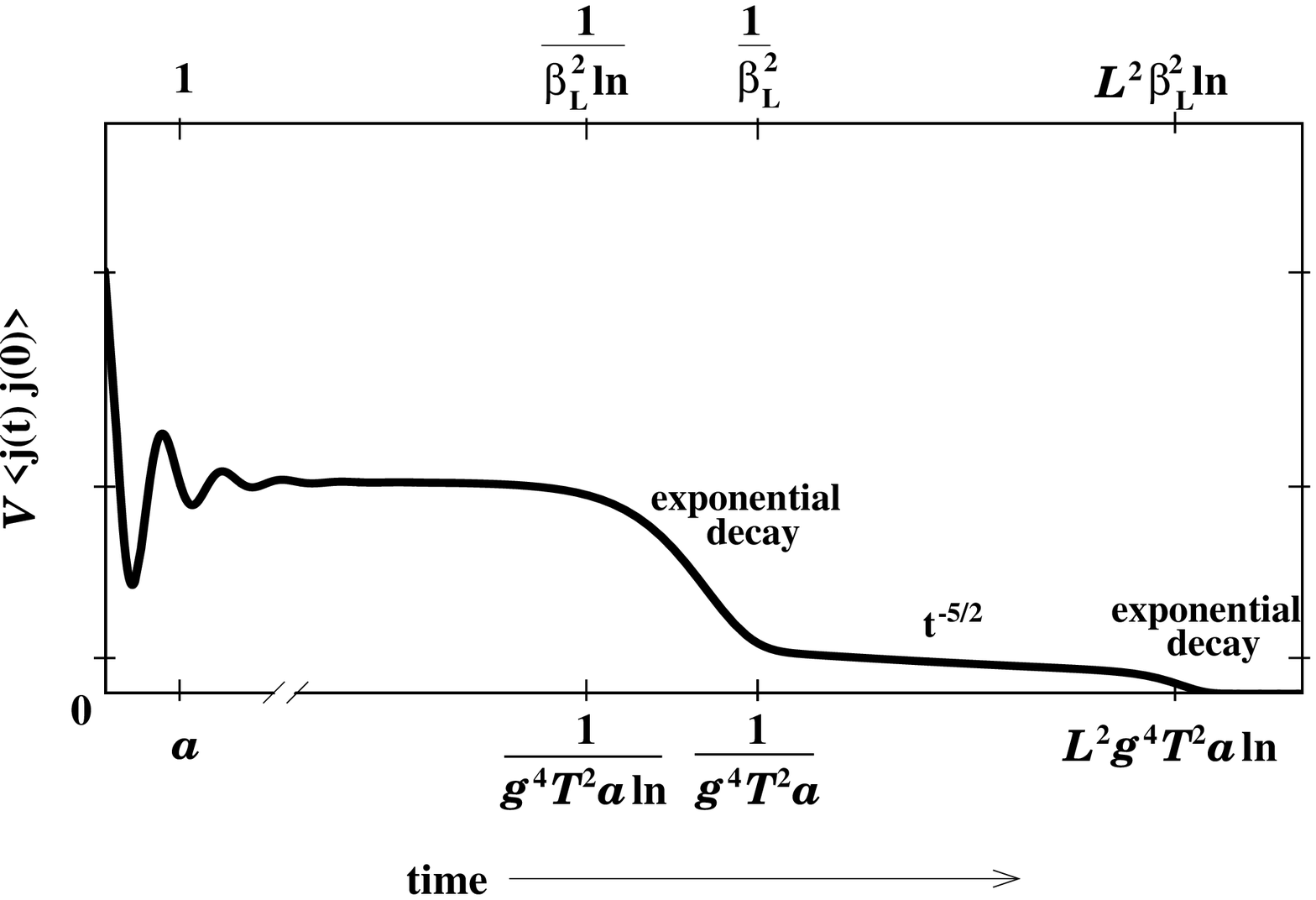}
   \end {center}
   \caption
       {%
         Same as fig.~\protect\ref{fig:lattice},
	 but in a large finite volume $L^3$ ($L \gg (g^4 T^2 a)^{-1}$).
       \label{fig:latticeVa}
       }%
   }%
\vspace*{0.3in}
\vbox
   {%
   \begin {center}
      \leavevmode
      
      \epsfbox [100 30 450 390] {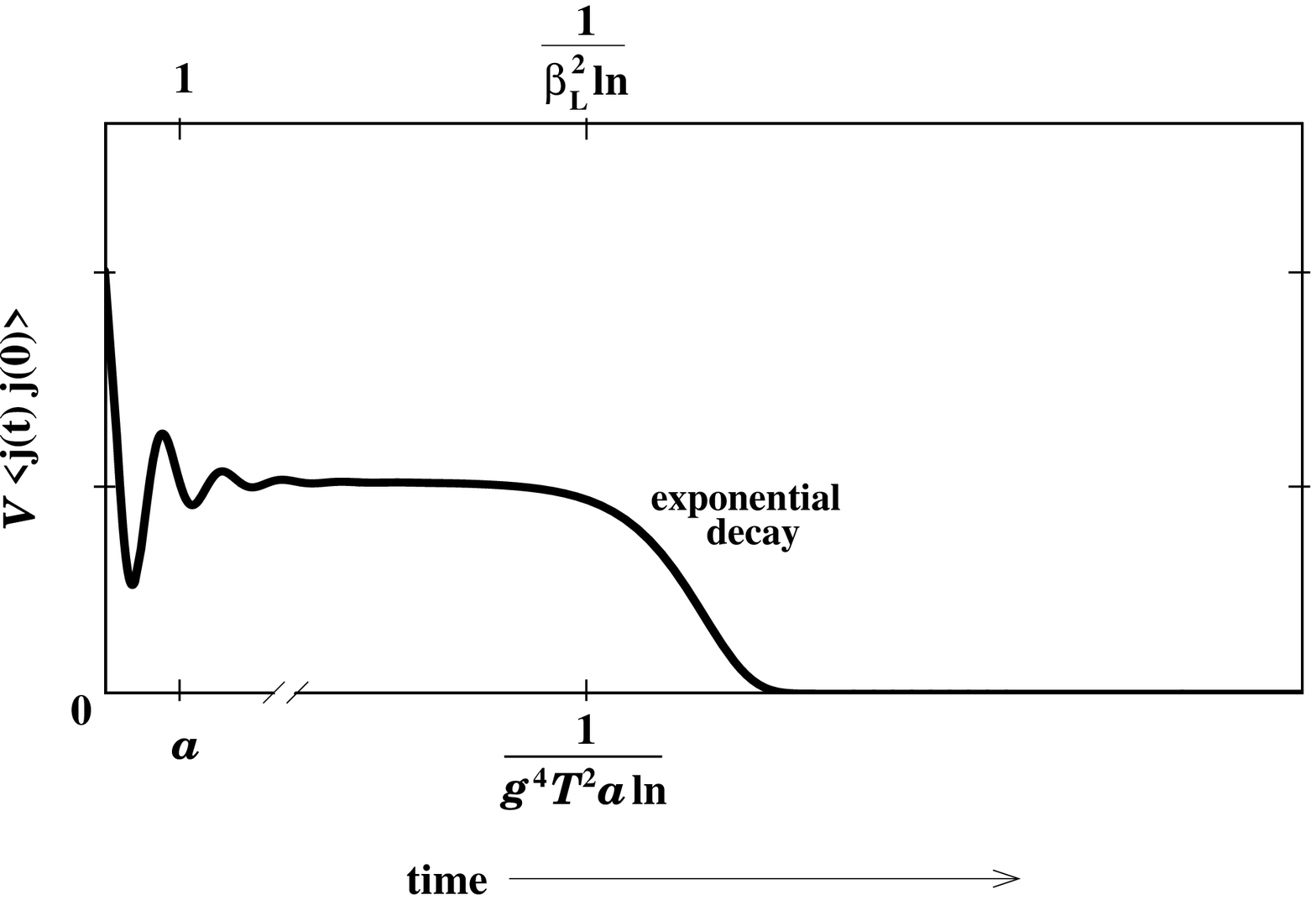}
   \end {center}
   \caption
       {%
	 Same as fig.~\protect\ref{fig:lattice},
	 but in a small volume $L^3$ 
	 ($L \protect\lesssim (g^4 T^2 a)^{-1}$).
         $\ln$ here means $\ln(\betaL)$ if $L \protect\gtrsim m^{-1}$
	 and $\ln(L/a)$ if $L \protect\lesssim m^{-1}$,
         where $m$ is the
         Debye mass $m \sim g T^{1/2} a^{-1/2} \sim \betaL^{-1/2} a^{-1}$.
	 (It is assumed that $L$ is still sufficiently large so that
	 the Poincar\'e recurrence time is far greater than the time scales
	 of interest.)
       \label{fig:latticeVb}
       }%
   }%
\end {figure}

First, consider the case where the volume is large compared to the
scale $\taubig$ for hydrodynamics.
In this case, the discussion of the last section still applies,
but now the largest fluctuation is limited to a size $L$,
where $L$ is the linear extent of the lattice.
The largest time scale for dissipation is then $t_L \sim L^2/\taubig$.
After this time, the correlation will again decay exponentially.
This can be seen more explicitly from (\ref{eq:lat tail}).
The integral over $\k$ is now a sum, and the smallest $k$ is
$2\pi/L$.  The very-very-large time behavior of the propagator is
then
\begin {eqnarray}
   V \langle{\j(t)\cdot\j(0)\rangle} &\simeq&
      {3\over L^3} \left(2\pi\over L\right)^2 \left[
      c_1^2 \, \chi_\epsilon \chi_q \,
             e^{-(D_\epsilon + D_q) (2\pi/L)^2 t}
      + c_2^2 \, \chi_q^2 
      \, e^{-2 D_q (2\pi/L)^2 t}
      \right]
\nonumber \\ &&
   \qquad\qquad\qquad\qquad\qquad\qquad\qquad\qquad\qquad\qquad\qquad
   (t \gg L^2/\taubig)
\end {eqnarray}
for an $L^3$ simple cubic lattice.
The resulting behavior for the correlator is depicted qualitatively
in fig.~\ref{fig:latticeVa}.

On the other hand, if the lattice is smaller than $\taubig$, then
the hydrodynamic power-law behavior will never develop.  In this
case, the correlator behaves as shown in fig.~\ref{fig:latticeVb}.


\section {What are classical simulations good for?}

    Real-time simulations of classical, thermal, lattice gauge field theory
are practical---one merely needs to integrate the classical equations
of motion forward in time for a sample of Boltzmann-distributed
initial conditions.
Unfortunately, no comparable practical, non-perturbative method
for studying real-time dynamics in thermal quantum field theories
is known.
For the specific purpose of computing the topological transition rate
in hot non-Abelian theories,
it was argued in ref.\ \cite {ambjorn} that the relevant degrees of freedom
are essentially classical, and therefore it should be possible
to determine the transition rate in the quantum theory by studying
a hot classical lattice theory.
In \cite {ASY} it was pointed out that the topological transition
rate is sensitive to the damping rate for gauge field fluctuations
with $O(g^2 T)$ momenta, and that this damping rate depends on
scattering off the thermally-distributed hard excitations.
In the quantum theory, the Bose distribution cuts off the
distribution of hard excitations at momenta of $O(T)$,
while a classical lattice theory instead has an $O(1/a)$ cut-off
from the finite lattice spacing.
Consequently, the damping rates for soft $O(g^2T)$ modes differ
between the quantum and classical theories.
Nevertheless, because these damping rates are perturbatively calculable
\cite {Arnold},
one can still, in principle, determine the quantum transition rate by
applying a calculable correction factor to the (lattice spacing dependent)
result of a classical lattice theory
\cite {kinetic swarms,moore}.%
\footnote
    {%
    We are glossing over the difficulties caused by the loss of
    rotational invariance in the classical lattice theory and
    resulting anisotropy in the classical damping rates
    \cite {Arnold,kinetic swarms}.
    However, this does not change the fact that essentially equivalent physics
    controls the dynamics of the relevant $p\sim g^2 T$, $\omega \sim g^4 T$
    fluctuations in the quantum and classical lattice theories.
    }

One might have hoped that thermal classical lattice theories,
and hot quantum theories, would generate the same behavior
(up to calculable correction factors) for many other observables
probing low-frequency, real-time response.
The example of the current-current correlator studied in this paper
shows that this cannot be universally valid.
At short times ($t \sim 1/T$ or $t \sim 1/a$, respectively)
the very different behavior of the current correlation
in the quantum and classical theories (illustrated in figs.~\ref{fig:summary}
and \ref {fig:lattice}) is surely no surprise;
at these times the correlation is sensitive to the precise
form of the cutoff in the distribution of hard excitations, which 
is completely different in the two theories.
However, the decay rate which characterizes the exponential regime
is also sensitive to scattering rates off hard excitations, and
no simple quantitative relation connects the resulting values
for the classical and quantum theories.
Finally, even the exponents of the long-time power-law tails differ
due to the very different nature of hydrodynamic fluctuations
in the quantum (continuum) and classical (lattice) theories.

Naturally, the same issues arise in the analysis of virtually any other
correlation function, although the details will differ.
For example, the correlator of the gauge-invariant scalar
operator $\int_\x \Phi^\dagger(\x) \Phi(\x)$,
also studied by Tang and Smit \cite{tang&smit},
has much smaller
short-time ($t\sim 1/a$)
oscillations in
the lattice theory because the one-loop perturbative
contribution to this correlator is IR dominated,
in contrast to the UV dominance of the current correlator.
As discussed by Tang and Smit,
the most noticeable feature is a decaying oscillation associated instead
with the thermal mass, at time scales $t \sim 1/m \sim 1/gT$.%
\footnote{
  Actually, the scaling of time scales is somewhat complicated in Tang
  and Smit's simulations, because they not only vary
  $\betaL \sim (g^2 T a)^{-1}$
  but vary the zero-temperature Higgs mass as well.  In this paper, we
  have routinely treated the zero-temperature Higgs mass as negligible.
}
Since $\Phi^\dagger \Phi$ has a non-vanishing equilibrium expectation
value, the two-point correlator of this operator does not decay to zero,
but will instead approach the square of the
(perturbatively computable) expectation value.
Just like the current-current correlator,
the time dependence of the $\Phi^\dagger \Phi$ correlator
will show a plateau in the free kinetic regime
followed (in infinite volume)
by initially exponential and later power-law
relaxation to the non-zero equilibrium value.
And, just as was shown for the current correlator,
quantitative aspects of this relaxation will differ between
the continuum quantum theory and the corresponding
classical thermal lattice theory.

Some readers acquainted with the role of classical lattice simulations in
studying the rate of high-temperature baryon number (B) violation may be
uneasy.  The time scale of B-violating processes is argued to be $t \sim 1/g^4
T$ at high temperature \cite{ASY}.  But that is the very time scale where, for
the case of current-current correlations, we've argued lattice simulations
begin to fail to even {\it qualitatively} reproduce the continuum answer.
There is,
however, a very important difference between these two cases.  At $t \sim
1/g^4 T$, the correlator of the space-averaged current is dominated by
collective fluctuations of hard modes ($p \sim T$), where the wavelength $L$
of the collective fluctuation is order $1/g^4 T$.  That is, large times
correspond to equally large distances, and one is on the edge of the
hydrodynamic regime ($t,L \gg (g^4 T)^{-1}$),
which the lattice fails to reproduce.
B violating processes, in contrast, are dominated by soft modes
($p \ll T$).
The soft modes of interest have momenta $p \sim g^2 T$, and
the fluctuations of interest have wavelength $L \sim 1/g^2 T$.
It is these soft modes
which should have an effective classical description,
and the physics is different than the hard-mode physics that dominates
hydrodynamic behavior.
Nonetheless, one might still fear trouble%
\footnote{
  That is, trouble more disastrous than what could be accommodated by some
  sort of perturbative matching.
}
because,
as argued in ref.~\cite{ASY}, the dynamics of the soft modes is
affected by hard modes with which they interact.  Does the lattice
do anything terrible to the dynamics of these hard particles, and hence
indirectly to the dynamics of the soft modes, on a time scale of $1/g^4 T$?
This question was addressed by Huet and Son in ref.~\cite{Huet&Son}.
The answer is
that it doesn't matter.  The time for a hard particle to fly across
the region of a B violating process is given by the size of that region: $L
\sim 1/g^2 T$.  What happens subsequently to the hard particle is irrelevant.
Since $1/g^2 T$ is small compared to $\taubig$, the hard-particle collisions
which produce hydrodynamic behavior are not relevant to the process of
B violation.


\acknowledgements

\noindent
This work was supported by the U.S. Department of Energy,
grant DE-FG03-96ER40956.


\appendix

\section {Fermionic charge carriers}
\label {app:fermions}

If the global charge associated with the current $\j$ is carried
by fermion fields, then the one-loop diagram fig.~\ref{fig:pert} gives
\begin {eqnarray}
   V \langle j_i(t) j_j(0) \rangle &=&
   \int_\p {1 \over 4 \omega_\p^2} \>
   {\rm tr}
   \left[\>
    \gamma_i
    \left(
	(1-n_{\omega_\p}) \, \Lambda^+_\p \, e^{-i \omega t}
	- n_{\omega_\p} \, \Lambda^-_{-\p} \, e^{i \omega t}
    \right)
    \right.
\nonumber
\\ && \>\quad\qquad {} \times
    \left.
    \gamma_j
    \left(
	(1-n_{\omega_\p}) \, \Lambda^-_{-\p} \, e^{-i \omega t}
	- n_{\omega_\p} \, \Lambda^+_\p \, e^{i \omega t}
    \right)
   \,\right]
   \\ &=&
   {\textstyle {2\over 3}} \,
   \delta_{ij} \,
   q^2
   \int_\p \> v_\p^2
   \left[ \,
    n_{\omega_\p} (1 - n_{\omega_\p})
    + n_{\omega_\p}^2 e^{2i\omega_\p t}
    + (1 - n_{\omega_\p})^2 e^{-2i\omega_\p t}
   \,\right]
\nonumber
\\ && {}
   + \delta_{ij} \,
   q^2
   \int_\p \> {m^2 \over \omega_\p^2}
   \left[ \,
    n_{\omega_\p}^2 e^{2i\omega_\p t}
    + (1 - n_{\omega_\p})^2 e^{-2i\omega_\p t}
   \,\right],
\label{eq:fermions}
\end {eqnarray}
where%
\footnote{
  For those who prefer the $+\,{-}\,{-}\,{-}$ metric convention over ${-}+++$,
  the $\gamma$ matrices would conventionally be normalized so that
  $\Lambda^\pm_\p \equiv \pm \rlap\slash p + m$.
}
$\Lambda^\pm_\p \equiv \mp i\rlap\slash p + m$
and $p^0 \equiv \omega_\p$.
$n_\omega = 1/(e^{\beta \omega} + 1)$ is the Fermi distribution.
As in the bosonic case, the prefactor $q^2$ represents the sum of
squared charges and,
because of the two spin states,
our convention is $q^2=2$ for the number current
of a single Dirac fermion.
The second term of the result (\ref {eq:fermions})
gives a subleading contribution (suppressed by $m/T$)
and will be neglected.
Separating the initial transient from the asymptotic constant,
as in (\ref {eq:CD}a), leads to
\begin {mathletters}%
\label {eq:CDf}%
\begin {eqnarray}
  C &=& {\textstyle {2\over3}} \int_\p v_\p^2 \;
  n_{\omega_\p} (1 - n_{\omega_\p})
  ,
\label{eq:Cf}
\\
  D(t) &=&
    {\textstyle {2\over3}}
    \int_\p v_\p^2
    \left[
      n_{\omega_\p}^2 e^{2i\omega_\p t}
      + (1 - n_{\omega_\p})^2 e^{-2i\omega_\p t}
    \right]
  .
\label{eq:Df}
\end {eqnarray}%
\end {mathletters}%
At $t=0$, one may easily see that ${\rm Re} \> D(0) = -2 C$.
Hence, the current-current correlation rises from $-C$,
crosses zero, and levels out at $C$.
For times large compared to $1/T$,
the qualitative behavior is identical to that shown
in fig.~1.
For massless continuum fermions, the integrals (\ref {eq:CDf})
may be evaluated analytically and yield
\begin {equation}
   C = {\textstyle {1\over 18}} T^3 , \quad
   D(t) = {{\textstyle {2\over 3}}T^3 \over \sinh(2\pi Tt)} \left[
       (2\pi Tt + i\pi) \left(
             \coth^2(2\pi Tt) - {\textstyle{1\over2}}\right)
	   - \coth(2\pi Tt)
   \right] .
\end {equation}


\section {Long-time decay of $D(t)$}
\label {app:oscillate}

    The long-time behavior of $D(t)$ may be extracted
by applying a stationary-phase approximation to the three-dimensional
momentum integral (\ref {eq:D}) defining $D(t)$,
or equivalently by analyzing the effect on the one-dimensional
Fourier transform (\ref{eq:fourier}) of the non-analyticities in the spectral
density $\rho_{vv}(\omega)$ (\ref{eq:rho}).
In the massive continuum theory, the only saddle point of $e^{2i \omega_\p t}$
occurs at $\p = 0$ and the resulting saddle point contribution
is
\begin {eqnarray}
    D(t)_{\rm cont} &\sim& {\textstyle {2\over 3}} \> {\rm Re}
    \int {d^3p \over (2\pi)^3} \>
    {p^2 \over m^2}  {T^2 \over m^2} \> e^{2 i [m + p^2/2m + O(p^4/m^3)] t}
\nonumber
\\
    &=&
    {T^2 \> t^{-5/2} \over (4\pi m)^{3/2}} \>
    \cos (2mt-{\textstyle {3\over 4}}\pi)
    ~ \left( 1 + O\left({1\over mt}\right) \right).
\label {eq:Dasym1}
\end {eqnarray}
At intermediate times, $\beta \ll t \ll m^{-1}$,
one may show that
\begin {equation}
    D(t)_{\rm cont} = D(t)_{\rm massless} ( 1 + O(m^2 \beta^2))
    - {T^2 m \over 4\pi} (1 + O(m^2 t^2)) \,,
\label {eq:Dasym2}
\end {equation}
where
$
    D(t)_{\rm massless} \sim {4\over 3} \, T^3 e^{-4\pi Tt}
    (2\pi t T - 1 + i \pi)
$
is the massless continuum result of (\ref{eq:CD_massless}).

For the lattice theory (on a simple cubic lattice),
the dispersion relation $\omega_\p$ has
additional saddle points at
$\p = (0,0,\pm\pi)$, $(0,\pm\pi, \pm\pi)$, and
$(\pm\pi, \pm\pi, \pm\pi)$,
on the edges and corners of the Brillouin zone,
with frequencies of $2/a$, $2\sqrt 2/a$, and $2\sqrt 3/a$,
respectively.
Each saddle point produces an oscillating $t^{-5/2}$
contribution at the associated frequency,
so that
\begin {eqnarray}
    D(t)_{\rm lattice} &\sim&
    D(t)_{\rm cont} +
    {T^2 \, a^{3/2} \over t^{5/2}}
    \left[
    {\cos (4t/a+{\textstyle {3\over 4}}\pi) \over (8\pi)^{3/2}}
    +
    {\cos (4\sqrt 2\,t/a-{\textstyle {3\over 4}}\pi)
          \over (8\sqrt 2\,\pi)^{3/2}}
    +
    {\cos (4\sqrt 3\,t/a+{\textstyle {3\over 4}}\pi)
          \over (8\sqrt 3\,\pi)^{3/2}}
    \right]
\nonumber
\\ && \hspace*{1.5in} \times
    \left( 1 + O\left({a/ t}\right) \right),
\label {eq:Dlat}
\end {eqnarray}
where $D(t)_{\rm cont}$ behaves as shown in (\ref {eq:Dasym1})
for $t \gg m^{-1}$,
or the second term of (\ref {eq:Dasym2}) for $a \ln \ll t \ll m^{-1}$.
Note that, for times not much larger than $a$,
the amplitudes of the lattice oscillations in (\ref {eq:Dlat})
are comparable to the $O(T^2 /a)$ steady-state value $C$
of the (perturbative result) for the current correlator.
At times of order $1/m$, where the continuum mass oscillations
begin to be evident, the amplitude of the oscillation has dropped
to $O(T^2 m)$, or $O(ma) = O(\betaL^{-1/2})$ times the steady-state value.


\end {document}